\documentclass[lettersize,journal]{IEEEtran}
\usepackage{cite}
\usepackage{caption} 
\usepackage{amsmath,amssymb,amsfonts}
\usepackage{algorithmic}
\usepackage{pdfpages}
\usepackage{graphicx}
\usepackage{booktabs}
\usepackage{textcomp}
\usepackage{xcolor}
\usepackage{tablefootnote}
\usepackage[misc]{ifsym}
\usepackage{subfig}
\usepackage{adjustbox}
\usepackage{amsmath,amssymb,amsfonts}
\usepackage{graphicx}
\usepackage{amsthm}
\usepackage{multirow}
\usepackage{bibentry}
\usepackage{listings} 
\usepackage{float}
\usepackage{color}
\usepackage{url}
\usepackage[ruled,linesnumbered]{algorithm2e}
\usepackage{makecell}
\usepackage{booktabs}

\begin{document}

\title{Split Fine-Tuning for Large Language Models in Wireless Networks}

\author{Songge~Zhang, 
        Guoliang~Cheng, 
        Xinyu~Huang,
        Zuguang Li, ~\IEEEmembership{Graduate Student Member,~IEEE,}
        Wen~Wu,~\IEEEmembership{Senior Member,~IEEE,}
    Lingyang~Song,~\IEEEmembership{Fellow,~IEEE,}
        and~Xuemin~(Sherman)~Shen,~\IEEEmembership{Fellow,~IEEE}
\thanks{Songge Zhang is with the School of Electronic and Computer Engineering, Peking University, Shenzhen, 518000, China, and also with the Frontier Research Center, Pengcheng Laboratory, Shenzhen, 518055, China (email: zhangsongge@stu.pku.edu.cn);

Guoliang Cheng, Zuguang Li, and Wen Wu are with the Frontier Research Center, Pengcheng Laboratory, Shenzhen, 518055, China (email: \{chenggl, lizg01, wuw02\}@pcl.ac.cn);

Lingyang Song is with the State Key Laboratory of Advanced Optical Communication Systems and Networks, School of Electronics, Peking University, Beijing, 100871, China, and also with the School of Electronic and Computer Engineering, Peking University Shenzhen Graduate School, Shenzhen, 518055, China (e-mail: lingyang.song@pku.edu.cn);

Xinyu Huang and Xuemin (Sherman) Shen are with the Department of Electrical and Computer Engineering, University of Waterloo, Waterloo, N2L 3G1, Canada (email: \{x357huan, sshen\}@uwaterloo.ca).}
}



\maketitle

\begin{abstract}
Fine-tuning is the process of adapting the pre-trained large language models (LLMs) for downstream tasks. Due to substantial parameters, fine-tuning LLMs on mobile devices demands considerable memory resources, and suffers from high communication overhead and long fine-tuning delay.
In this paper, we propose an efficient LLM fine-tuning scheme in wireless networks, named \underline{S}plit \underline{F}ine-\underline{T}uning (SFT), which can accommodate LLM fine-tuning on mobile devices. Specifically, an LLM is split into a server-side part on the edge server and a device-side part on the mobile device to satisfy the device-side memory constraint. All devices share a server-side model and perform parallel fine-tuning to reduce fine-tuning delay. In addition, to reduce significant communication overhead incurred by data exchange between devices and the edge server, we propose a data compression scheme by jointly leveraging sparsification, stochastic quantization, and lossless encoding methods. 
Furthermore, we formulate a fine-tuning delay minimization problem under accuracy and memory constraints, taking device heterogeneity and channel dynamics into account.  
To solve the problem, the nonlinear mixed-integer problem is decoupled into two subproblems in different timescales.
The two-timescale resource management algorithm is proposed to jointly optimize the compression rate and transformer block allocation in the large timescale using the augmented Lagrangian method, and determine spectrum resource allocation in the small timescale via sequential quadratic programming.
Extensive simulation results demonstrate that the proposed scheme can reduce the fine-tuning delay by up to 80.2\% and communication overhead by 93.6\% compared to state-of-the-art benchmarks, while satisfying device-side memory and model accuracy constraints.

\end{abstract}


\section{Introduction}
Recent advancements in large language models (LLMs), such as ChatGPT, LLaMA, and Vision Transformer~\cite{NLP,ShiwenMao,XiaopingZhang2}, have sparked a new wave in artificial intelligence (AI) by showcasing impressive abilities such as improved generalization and inference. These advancements have facilitated extensive applications across various fields, such as chatbots, search engines, writing assistants, and multimodal systems~\cite{application}, illuminating the vision of artificial general intelligence~\cite{COMST,ZhangJun}. Deploying large models on mobile devices enables timely task responses near the mobile devices in wireless networks.

Fine-tuning is a critical technique for LLM to support personalized services and downstream tasks~\cite{VehicleNetwork}.
Fine-tuning adjusts LLM parameters using local data for specific tasks, retaining pre-trained performance. 
Fine-tuning can be performed through full parameter tuning or parameter-efficient fine-tuning (PEFT). 
PEFT updates only a small subset of parameters or additional modules while keeping the rest of the model frozen, which can reduce the computation workload on devices~\cite{PEFT2}.
In wireless networks, LLMs are fine-tuned based on a number of mobile devices' data, such as behavioral records, location information, and multimedia content, which cannot be uploaded to the centralized cloud due to data privacy concerns~\cite{OJVT}. Therefore, fine-tuning should be performed in a distributed manner via the collaboration among multiple devices without sharing raw local data~\cite{LiXiangyang}.


\begin{table*}[t]
    \centering
    \caption{Comparison between Typical Mobile Device Resources and Memory Requirements for Deploying Typical LLMs.}
    \label{FirstTable}
    \begin{tabular}{|c|c||c|c|c|}
        \hline
        \multicolumn{2}{|c||}{Resources of Typical Mobile Devices} &
        \multicolumn{3}{c|}{Required Resources for Deploying Typical LLMs} \\ \hline
        Devices & Memory & Models & \# Parameter & Memory \\ \hline
        Raspberry Pi-4B & 4 GB & LLaMA-7B & 7 Billion & 28 GB \\ \hline
        NVIDIA Jetson Nano & 8 GB & LLaMA-65B & 65 Billion & 260 GB \\ \hline
        NVIDIA Jetson TX2 & 8 GB & GPT-3 & 175 Billion & 700 GB \\ \hline
        NVIDIA Jetson Xavier NX & 16 GB & PaLM & 540 Billion & 2.15 TB \\ \hline
    \end{tabular}
\end{table*}
In the literature, fine-tuning explores a distributed paradigm with PEFT techniques such as low-rank adaptation (LoRA)~\cite{XiaopingZhang}. LoRA can modify a small portion of the LLM parameters instead of the entire LLM in the fine-tuning process. Although distributed schemes with PEFT are highly parameter-efficient, they still demand substantial memory to deploy the entire LLM, exceeding the memory capacity of typical mobile devices~\cite{WirelessCom.,JSTSP3}. As shown in Table~\ref{FirstTable}, typical NVIDIA Jetson mobile devices cannot meet the memory requirements for training LLMs like GPT-3 and LLaMA~\cite{ShiwenMao}. To address this issue, split learning (SL) is a potential solution, which can divide LLMs into server-side and device-side parts, offloading most of the model to the server, thereby satisfying the device's memory constraints~\cite{JSTSP2}.

However, designing an efficient SL scheme for LLMs fine-tuning presents several challenges. Firstly, in the vanilla SL, all devices interact with the server in a sequential manner to complete the fine-tuning process, which results in a long fine-tuning delay. Secondly, the SL scheme needs to periodically transmit intermediate activations and activation gradients between the server and devices, resulting in significant communication overhead. The activation and gradient values are typically large in the LLM. 
Considering ViT-16 and a mobile device with 5,000 data samples and 196 tokens, the total immediate activation data volume at the cut layer can be approximately 2.81 GB for one fine-tuning epoch~\cite{VIT16}. Thirdly,  heterogeneous devices' computational capabilities and dynamic channel conditions result in varying delay for each device, known as the straggler effect, which increases the fine-tuning delay.


To address the above, we propose a \underline{s}plit \underline{f}ine-\underline{t}uning (SFT) scheme that satisfies memory constraints by splitting the LLMs into device and server parts. All devices collaboratively fine-tune LLMs with the server using the devices' local data and parallelly update local device-side LoRAs, which are aggregated in each epoch. 
The fine-tuning process on each device is conducted in parallel, and the server executes all devices' training processes with only one server-side model.
Once all devices complete their fine-tuning, each device uploads the updated device-side LoRAs for aggregation.
The proposed scheme can efficiently reduce memory consumption and computation workload on the device by splitting the LLM. 
In addition, we design a novel compression scheme and resource management algorithm.
Specifically, the compression scheme includes Top-K sparsification, stochastic quantization, and lossless encoding, which focus on the most significant parameter for the accuracy performance of LLMs and reduce the data volume within an allowable accuracy degradation.

Additionally, we analyze the delay performance, communication and computation overhead, as well as memory consumption of the proposed scheme.
Furthermore, we formulate a fine-tuning delay minimization problem under constraints of device-side memory and accuracy, which is a nonlinear mixed-integer programming problem.
We decompose the problem into two subproblems by distinguishing different variables in the different timescales: a large-timescale subproblem for determining compression rates and block allocation, and a small-timescale subproblem for determining the device's bandwidth allocation. Specifically, we use the augmented Lagrangian method to iteratively solve the former subproblem, addressing its non-convex constraints, and employ sequential quadratic programming (SQP) to iteratively approximate the latter nonlinear subproblem as a series of quadratic subproblems. \textcolor{black}{The extensive simulation results demonstrate that the proposed scheme can reduce total fine-tuning delay up to $80.2\%$ and communication overhead up to $93.6\%$ compared to the state-of-the-art scheme.}
The main contributions of this paper are summarized as follows:
\begin{enumerate}
    \item We propose an SFT scheme for LLM fine-tuning in wireless networks, enabling collaborative and parallel fine-tuning across multiple devices.
    \item We propose a novel compression scheme that significantly reduces communication overhead.
    \item We analyze the performance of the SFT scheme, including fine-tuning delay, memory consumption, communication overhead, and computational workload.
    \item We propose a resource management algorithm to reduce fine-tuning delay by optimizing compression rates, transformer block allocation, and spectrum resources.
\end{enumerate}

{\color{black}{The remainder of this paper is organized as follows. Section~\ref{2} reviews the related works. Section~\ref{3} presents the system model. Section~\ref{4} introduces the proposed scheme. Section~\ref{5.5} introduces the performance analysis. Sections~\ref{5} and ~\ref{6} detail the problem formulation and the corresponding solution, respectively. Section~\ref{7} provides the simulation results, and finally Section~\ref{8} concludes the paper.}}

\section{Related Work}\label{2}
A wide range of studies aim to improve LLM fine-tuning performance from different perspectives. Adapters achieve parameter-efficient fine-tuning (PEFT) by introducing lightweight trainable modules between existing layers of the model while freezing the rest~\cite{PEFT1}. Prefix-tuning optimizes a small set of prefix vectors prepended to the input at each layer, leaving the model's original parameters unchanged~\cite{PEFT3}. LoRA reduces the parameter update size by injecting low-rank trainable matrices into the attention mechanism while keeping the majority of the model frozen~\cite{PEFT5}.
To achieve collaborative training over wireless networks, some work leverages FL to allow multiple users to collaboratively fine-tune LLMs without the need to share data~\cite{FD3}. In these studies, the entire model is deployed on local devices for training, with model parameters or gradients aggregated at each round~\cite{SHEN6}. 
Some studies explore FL with PEFT techniques. \cite{FD4} propose the PromptFL framework, which enables federated participants to train shared prompts instead of the entire model. Jiang \emph{et al.}~\cite{FD6} introduce a soft label-enhanced federated fine-tuning approach that incorporates LoRA to reduce computational and communication costs. Cai \emph{et al.}~\cite{FD7} propose FedAdapter, which progressively adjusts adapter configurations to identify the most efficient settings, accelerating model convergence.
Additionally, some studies enable LLM fine-tuning in FL using techniques like distillation, pruning, and quantization. Liu \emph{et al.}~\cite{FD5} introduce an adaptive quantization scheme with ensemble distillation, which compresses a large model into a smaller one and applies cluster partitioning for heterogeneous model training. Greidi \emph{et al.}~\cite{JSTSP4} propose sparse training to reduce computation and communication costs in FL.
These studies enable collaborative LLM fine-tuning across multiple devices while reducing the computational load on devices and communication overhead.

\begin{figure*}[t]
\centering
\subfloat[Proposed SFT]{\includegraphics[width=3.0in]{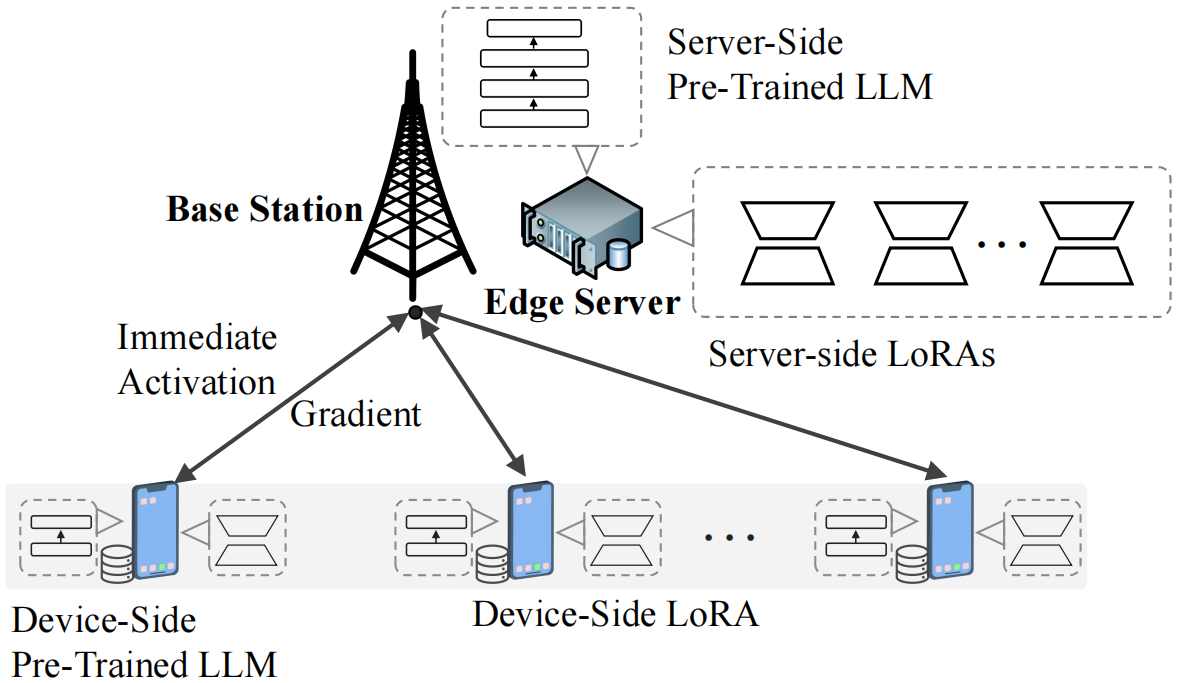}\label{SplitLLM}}\hfil
\subfloat[Transformer Block]{\includegraphics[width=2.9in]{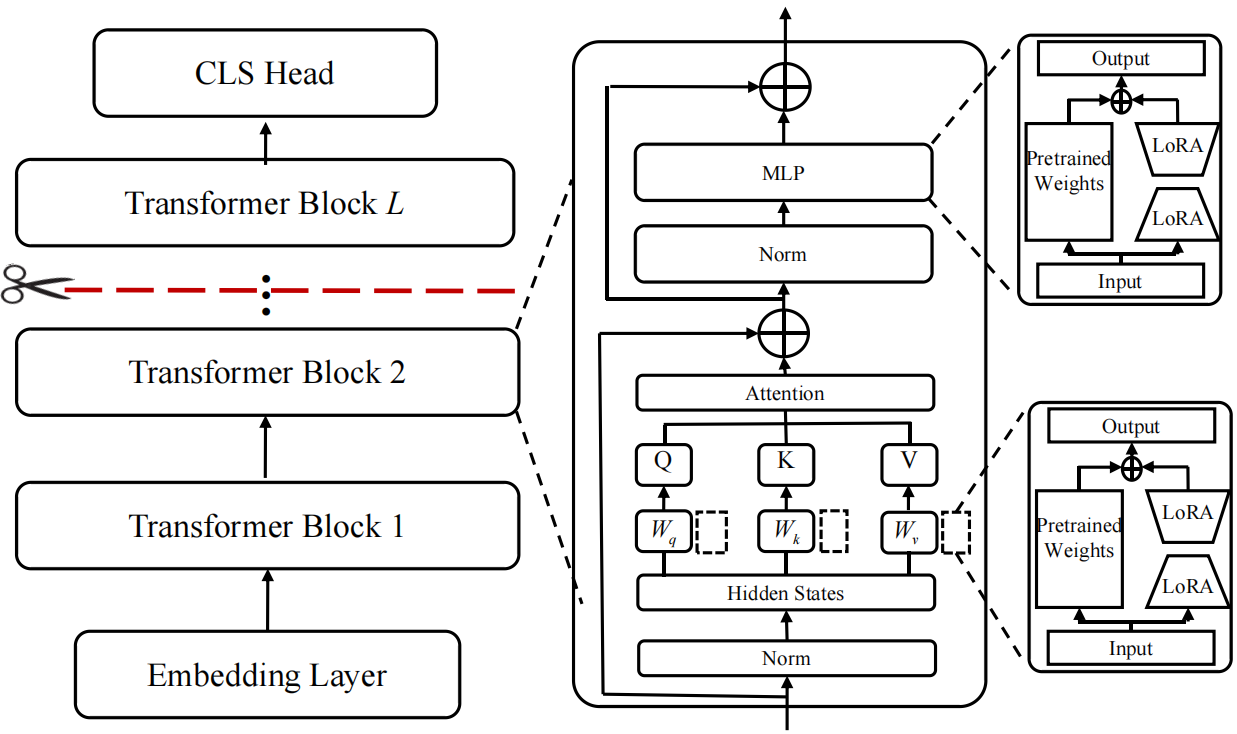}\label{CutLayer}}
\caption{(a) In the SFT frame, devices are trained parallelly with a shared server-side pre-trained model and multiple LoRAs$\mathrm{;}$ (b) All Transformer blocks are divided into the server-side and the device-side parts. Each transformer block consists of an MSA and an MLP, both of which are composed of pre-trained model weights and corresponding LoRA.}
\label{fig_cnn_performance}
\end{figure*}

In addition to FL, SL is another potential solution for enabling fast model training. The SL scheme can reduce the computational workload and memory consumption on devices by partitioning the model~\cite{Songge1}. To accelerate the training of AI models, Wu \emph{et al.} introduced a parallel-then-sequential SL strategy to enhance training speed~\cite{SL2}. Kim \emph{et al.} and Xu \emph{et al.} proposed a synchronized SL (SFL) scheme, allowing multiple server-side and device-side models to train simultaneously, thereby speeding up the SL process~\cite{SL4}. Liao \emph{et al.} addressed system heterogeneity and accelerated model training by optimizing bandwidth allocation~\cite{SL5}.To more efficiently improve SL, joint optimization of model partitioning and resource management has also been explored. For example, in~\cite{SL6}, the optimization of the partitioning layer and bandwidth allocation was proposed to mitigate the straggler effect in wireless SFL.
Very recently, a few pioneering works have been devoted to enhancing LLM fine-tuning via SL. 
Wu \emph{et al.} split the large model into different components, deploy them across multiple devices, and fine-tune these components on the device side using a pipeline approach~\cite{TWCMagazine}.
Chen \emph{et al.} split the Adapter from the large model's backbone and deploy it on devices for fine-tuning~\cite{ChenXu}.
Different from the existing scheme,  we design a parallelized split scheme to accelerate the delay of LLM fine-tuning. Additionally, we implement compression strategies to reduce the communication overhead between the edge server and mobile devices.

\section{System Model}\label{3}
\subsection{Considered Scenario}
In this paper, we consider a typical wireless network scenario consisting of multiple mobile devices and a base station equipped with an edge server with powerful computational and memory resource.
We define the set of devices as $ \mathcal{N}=\{1,2, \dots, N\}$. 
Each device possesses local data and operates a small portion of the LLM. The data from all devices contribute to fine-tuning tasks such as image classification, natural language processing, etc. The edge server is primarily responsible for the LLMs in a fine-tuning training task and also manages model aggregation at each training round.

In this paper, we consider a transformer-based LLM architecture consisting of $L$ layers. These transformer blocks are split into the server and devices, with the device locally processing $l$ blocks while the remaining $L-l$ blocks are possessed by the server. Forward and backward propagation between the edge server and devices is carried out through the transmission of intermediate activations and activation gradients. The parameter size of each block is related to the dimensions of the multi-head self-attention (MSA) and multi-layer perceptron (MLP).

\subsection{Fine-Tuning Model}
We focus on supervised fine-tuning tasks, utilizing the LoRA adapter to reduce the cost and parameter number of fine-tuning. We denote dataset of each device as \(\mathrm{D}_n = \{\mathbf{x}_j, \mathbf{y}_j\}_{j=1}^{|\mathrm{D}_n|}\), where \(|\mathrm{D}_n|\) denotes the total number of samples for device \(n \in \mathcal{N}\) and $\sum_{n \in \mathcal{N}} \mathrm{D}_n = \mathrm{D}$ where $\mathrm{D}$ represents the total number of samples across all devices. Here, \(\mathbf{x}_j\) is the \(j\)-th input sample, and \(\mathbf{y}_j\) is its corresponding label.
Let \(\mathbf{\Theta}\) represent the LLM parameters. 
The loss function for each sample, denoted as \(f_j(\mathbf{\Theta})\), measures the prediction error for the \(j\)-th sample. The objective of the training process is to minimize the empirical loss \(F(\mathbf{\Theta})\), which is defined as
\begin{equation}
   F(\mathbf{\Theta}) = \frac{1}{|\mathrm{D}|} \sum_{j=1}^{|\mathrm{D}|} f(\mathbf{x}_j, \mathbf{y}_j; \mathbf{\Theta}) = \frac{1}{|\mathrm{D}|} \sum_{j=1}^{|\mathrm{D}|} f_j(\mathbf{\Theta}). 
\end{equation}
This process seeks to minimize the average prediction error across all samples and identify the optimal parameter $\mathbf{\Theta}^\star$.

To enhance the efficiency of local model transmission and aggregation, we integrate a small, low-rank, homogeneous adapter into the locally pre-trained model.
As depicted in Fig. \ref{CutLayer}, the LoRA adapter is designed to retain the same input dimension \(d\) and output dimension \(h\) as pre-trained large models, and is applicable to linear, embedding, and convolutional layers. Particularly for linear layers, the adapter reduces the parameter number by decomposing the conventional parameter matrix \( \mathbb{R}^{d  \times h} \) into two smaller matrices \( \mathbf{A}\in\mathbb{R}^{d \times r} \) and \( \mathbf{B}\in\mathbb{R}^{r \times h} \), where the rank \( r \) is significantly lower than both \( d \) and \( h \). During the initialization phase, matrix \( \mathbf{A} \) is initialized with a Gaussian distribution with mean zero and variance \( \sigma^2 \), while matrix \( \mathbf{B} \) is set to zero~\cite{sun2024improving}. 
 In fine-tuning training, we define \(\mathbf{\Theta} \in \mathbb{R}^{d \times m}\) as a trainable weight matrix of the pre-trained model, where the corresponding model update is expressed as \(\mathbf{\Theta} + \Delta \mathbf{\Theta} = \mathbf{\Theta} + \mathbf{AB}\). 
 LoRA adapter configuration allows the parameter number to be reduced significantly compared to full-parameter fine-tuning, thereby enhancing computational and communication efficiency. Accordingly, we deploy the LoRA adapter within the considered scheme.
Thus, the objective of fine-tuning can be rewritten as 
\begin{equation}\label{loss function}
    \begin{aligned}
        F(w) = \frac{1}{|D|} \sum_{j=1}^{|D|} f_j(\mathbf{\Theta} + \Delta\mathbf{\Theta}).
    \end{aligned}
\end{equation}
which is to minimize the average prediction error across all samples and identify optimal parameter $\Delta\mathbf{\Theta}^\star$.

\begin{algorithm}[t]
\SetAlgoLined

Initialize the cut layer and LoRA adapter parameter;\\
Base station broadcasts $1$ to $l$ transformer block;\\
\For{training round $t = 1,2,\ldots,T$}{
    Base station broadcasts the latest device-side LoRA adapter;\\
        \For{device $n \in \mathcal{N}$}{
            \For{local epoch $k=1,2,\dots,K$}{
                \#  Device executes\\
                Randomly selects a mini-batch sample from the local data;\\
                Execute the FP of the device-side pre-trained model and LoRA adapter, then obtain an immediate result;\\
                Transmit immediate result to the server based on the compression scheme;\\
                \#  Server executes\\
                {
                    Receive the immediate result from the device;\\
                    Execute the FP of server-side pre-trained model and $n$-th LoRA adapter;\\
                    Update the server-side LoRA adapter;\\
                    Transmit the device-side LoRA adapter's gradient;\\
                }
            
            }
            Update the device-side LoRA adapter to the edge server;\\
        }
        All devices send the latest device-side LoRA adapter to the edge server;\\

Edge server aggregates server-side LoRA adapters and device-side LoRA adapters into a new LoRA.\\
}
\caption{Split fine-tuning scheme.}
\label{Split algorithm}
\end{algorithm}

\section{SFT Scheme}\label{4}

\subsection{SFT Scheme}
To enable collaborative fine-tuning on memory-constrained devices, we propose an innovative split fine-tuning architecture called SFT. The process of this scheme is detailed in Alg.~\ref{Split algorithm} and primarily includes the following steps.

\textbf{Initial stage}.
The initial stage of fine-tuning involves splitting the transformer blocks and distributing them to the devices.
In the initial phase, the pre-trained model, which consists of $L$ layers, is split into two parts: $\mathbf{\Theta}_u$ for the device-side model, and $\mathbf{\Theta}_s$ for the server model. 
The segmented pre-trained model is distributed from the edge server to the device only once throughout the training process, because the pre-trained model does not require updates and aggregation in each fine-tuning round.
We assume there exist $L$ LoRA adapters, denoted as $\mathcal{L}=\{1,\dots,L\}$.
Specifically, the device model consists of the $l$-th layer, denoted as $\mathcal{L}_u=\{1,2,...,l\}$. The server handles the remaining layers from $l+1$ to $L$, indicated as $\mathcal{L}_s=\{l+1,\dots,L\}$.
We assume the device-side LoRA adapter, $\Delta \mathbf{\Theta}_u$, consists of $\{\mathbf{A}^1, \mathbf{B}^1, \ldots, \mathbf{A}^{l}, \mathbf{B}^{l}\}$; and the server's adapter, $\Delta \mathbf{\Theta}_s$, comprises $\{\mathbf{A}^{l+1}, \mathbf{B}^{l+1}, \ldots, \mathbf{A}^L, \mathbf{B}^L\}$. These LoRA adapters are updated in each training round and distributed to the devices by the server.

\textbf{Forward propagation (FP) in device-side and server-side model}.
Assume the fine-tuning spans $T$ rounds. In round $t$ and $k$ epoch, each device randomly selects a mini-batch from their local data, denoted by $\mathcal{B}_{n}(t,k) \subseteq \mathrm{D}_{n}$, to perform FP. Here, $B=|\mathcal{B}_{n}(t,k)|$ represents the size of each mini-batch, and $\mathrm{D}_{n}$ is the local dataset of device $n$. The FP involves passing $\mathcal{B}_{n}(t,k)$ through $\mathbf{\Theta}_u(t,k)$ and $\Delta \mathbf{\Theta}_u(t,k)$ to produce the output $\mathbf{y}_{n}^u(t,k)$. This process can be mathematically represented~as
\begin{equation}
\mathbf{y}_{n}^u(t,k) = f(\mathbf{\Theta}_u(t,k), \Delta \mathbf{\Theta}_u(t,k), \mathcal{B}_{n}(t,k)).
\end{equation}
All devices transmit their locally computed outputs to the server. 
The server then conducts FP based on the server-side  $\mathbf{\Theta}_s(t,k)$ and $n$-th LoRA adapter $\Delta \mathbf{\Theta}_s(t,k)$. This process can be mathematically represented as
\begin{equation}
\mathbf{y}^s(t,k) = f(\mathbf{\Theta}_s(t,k), \Delta \mathbf{\Theta}_s(t,k), \mathbf{y}_{n}^u(t,k)).
\end{equation}
Since we consider only one pre-trained model on the server side, the server will perform the LoRA FP sequentially according to the order in which the immediate activation is received.

\textbf{Backward propagation (BP) in device-side and server-side model}.
BP requires updating the LoRA adapter parameters of the server and device.
The purpose is to minimize the global loss function as defined in Eq. \eqref{loss function}. The edge server calculates gradients for both the server-side and device-side LoRA adapter $\mathbf{g}_{n}^{u}(t,k)$ and $\mathbf{g}_{n}^{s}(t,k)$ based on the prediction outcomes $\mathbf{y}^s(t,k)$ and labels. Subsequently, the edge server updates server-side LoRA adapters parameters as follows:
\begin{equation}
    \Delta\mathbf{\Theta}_{s}^n(t,k) \leftarrow \Delta\mathbf{\Theta}_{s}^n(t-1,k) - \epsilon_s \mathbf{g}_{n}^{s}(t,k), \forall n \in \mathcal{N}.
\end{equation}
where $\epsilon_s$ denotes the learning rate for the LoRA adapter on the edge server. For simplicity, assume each transformer block is linked sequentially, with the LoRA adapter updating from the last to the cut $(l+1)$-th layer. When the gradient is updated to the cut layer, gradients for layers $1$ through $l$ of LoRA adapters are transmitted back to the corresponding device. The device then update the device-side LoRA adapters as follows:
\begin{equation}
    \Delta\mathbf{\Theta}_{u}^n(t) \leftarrow \Delta\mathbf{\Theta}_{u}^n(t-1,k) - \epsilon_u \mathbf{g}_{n}^{u}(t,k), \forall n \in \mathcal{N},
\end{equation}

\textbf{LoRA aggregation in each round}.
Each device will iteratively perform the first three stages for $K$ epochs. All devices will then upload their updated LoRA adapters to the edge server for aggregation. 
The server aggregates server-side and device-side LoRA adapters using the FedAvg method. The aggregation process can be mathematically represented as
\begin{equation}
    \Delta\mathbf{\Theta}_n^s(t+1,1) = \sum_{n=1}^N  \frac{D_{n}}{D} \Delta\mathbf{\Theta}_n^s(t,K),~ \forall n \in \mathcal{N},
\end{equation}
and 
\begin{equation}
      \Delta\mathbf{\Theta}_n^u(t+1,1) = \sum_{n=1}^N  \frac{D_{n}}{D}   \Delta\mathbf{\Theta}_n^u(t,K),~ \forall n \in \mathcal{N}.
\end{equation}
The aggregated process will involve a new fine-tuning round until the target accuracy is reached.

\subsection{Compression Scheme}
\begin{figure}[!t]
\centering
\includegraphics[width=3.0in]{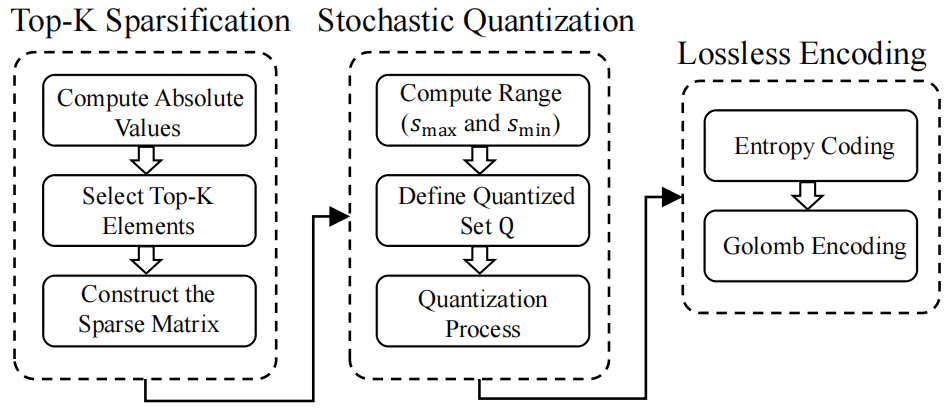}
\caption{Transmission compression scheme.}
\label{Sparsification}
\end{figure}

In the LLMs, the transmission of intermediate parameters between device-side transformer blocks and server-side transformer blocks can become a communication bottleneck. To address the high communication cost caused by transmitting the full activation matrix \( \mathbf{s}_{l} \) between transformer blocks, sparsification, quantization, and encoding techniques can be applied to reduce the amount of data transferred while retaining the most critical information~\cite{Guoliang2}. 
 Let \(\mathbf{s}_{l} \in \mathbb{R}^{N \times  D} \) represent the activations that are transmitted from the \( l \)-th transformer block to the \( (l+1) \)-th transformer block, where \( N \) is the number of patches (or tokens), and \( D \) is the embedding dimension. Directly transmitting the full matrix \( \mathbf{s}_{l} \) between blocks incurs a high communication cost, especially when \( N \) and \( D \) are large.
The transmission compression scheme can be represented as shown in Fig.~\ref{Sparsification}.

\textbf{Top-K sparsification}:
In Top-K sparsification, we selectively retain only the \(K\) largest-magnitude elements in the matrix \(\mathbf{s}_{l}\), while setting the remaining elements to zero. This approach effectively reduces the communication cost by focusing on the most significant activations. The steps for applying Top-K sparsification to \(\mathbf{s}_{l}\) are as follows:

\begin{enumerate}
    \item Compute absolute values: For each element \(\mathbf{s}_{l, i, j}\) in the matrix \(\mathbf{s}_{l}\), compute the absolute value \(|\mathbf{s}_{l, i, j}|\) to capture the magnitude of each activation.

    \item Select Top-K elements: Identify the indices of the \(K\) largest values based on their magnitudes. Let \(\mathcal{I}_K\) represent the set of indices corresponding to these top \(K\) elements.

    \item Construct the sparse matrix: Create a sparse version of \(\mathbf{s}_{l}\), denoted as \(\hat{\mathbf{s}}\), where only the elements in \(\mathcal{I}_K\) retain their original values, and all other elements are set to zero. This sparse matrix can be expressed mathematically as follows:
\begin{equation}
\hat{\mathbf{s}}_{l,i,j}= 
    \begin{cases} 
       \mathbf{s}_{l, i, j}, & \text{if } (i, j) \in \mathcal{I}_K; \\
       0, & \text{otherwise}.
    \end{cases}
\end{equation}
\end{enumerate}

By applying this process, we obtain a sparse matrix \(\hat{\mathbf{s}}_l\) that reduces the communication load between transformer blocks while preserving the most significant activation values. 
The sparsification rate, \(\rho\), is defined as the ratio of the number of retained elements \(K\) to the total number of elements in \(\mathbf{s}_{l}\), expressed as:
\begin{equation}
\rho = \frac{K}{\text{dim}(\mathbf{s}_{l})},
\end{equation}
where \(\text{dim}(\mathbf{s}_{l})\) denotes the total number of elements in the matrix \(\mathbf{s}_{l}\). The sparsification rate satisfies the condition:~
$\rho \in \mathbb{R}^+$.

\textbf{Stochastic quantization}:
Our aim is to utilize fewer bits for representing the non-zero parameter \( \hat{s}_{l} \) in the quantized vector. Initially, we compute the absolute values of \( \hat{s}_{l} \), denoted as \( | \hat{s}_{l} | \). We then identify the maximum and minimum non-zero values of \( | \hat{s}_{l} | \) within the set, expressed as \( s_{\text{max}} = \max \{|s|, s \in \hat{s}_{l} \}\) and \( s_{\text{min}} = \min \{|s|, s \in \hat{s}_{l} \} \), respectively. For a quantization level \( E \in \mathbb{Z}^+\), we define the set of quantized values \( Q = \{ Q_1, Q_2, \dots, Q_E \} \), where each quantization point \( Q_e \) is calculated as

\[
Q_e = \frac{e (s_{\text{max}} - s_{\text{min}})}{E} + s_{\text{min}}.
\]

Our objective is to map each non-zero scalar \( s \in \hat{s}_{l} \) into this discrete quantized set \( Q \). To achieve this, we select an index \( e \) such that \( Q_e \leq |s| \leq Q_{e+1} \), thereby determining the quantization interval for \( s \) as \( [Q_e, Q_{e+1}] \). The quantization procedure for any non-zero scalar \( s \in \hat{s}_{l} \) is then defined as

\[
\text{quant}(s, Q) = 
\begin{cases} 
\text{sgn}(s) \cdot Q_e, & \text{with probability } \frac{Q_{e+1} - |s|}{Q_{e+1} - Q_e}; \\
\text{sgn}(s) \cdot Q_{e+1}, & \text{otherwise},
\end{cases}
\]
where \( \text{sgn}(s) \in \{-1, +1\} \) determines the sign of \( s \). Then, this approach yields the sparse and quantized model update \( \text{quant}(\hat{s}_{l}, Q) \).

\textbf{Lossless encoding}:
Given the distribution characteristics of \( L_{t,i} \), where smaller indices tend to appear more frequently, entropy coding can be utilized to reduce the data size. Additionally, the sparse binary matrix \( m_{t,i} \) can be efficiently compressed using Golomb encoding~\cite{LossCode1,LossCode2}.

{\color{black}{Based on the established compression scheme, we can systematically adjust the combinations of \(\{ \rho, E \}\) and document the corresponding compression rates. Using this data, we then construct a piecewise linear function to predict the optimal compression strategy \(\{ \rho, E \}\) based on a given \( \beta \). This function can be effectively approximated by the server using a minimal set of public training data in an offline setting.}}

\subsection{Accuracy Performance}
\begin{figure}[!t]
\centering
\includegraphics[width=2.3in]{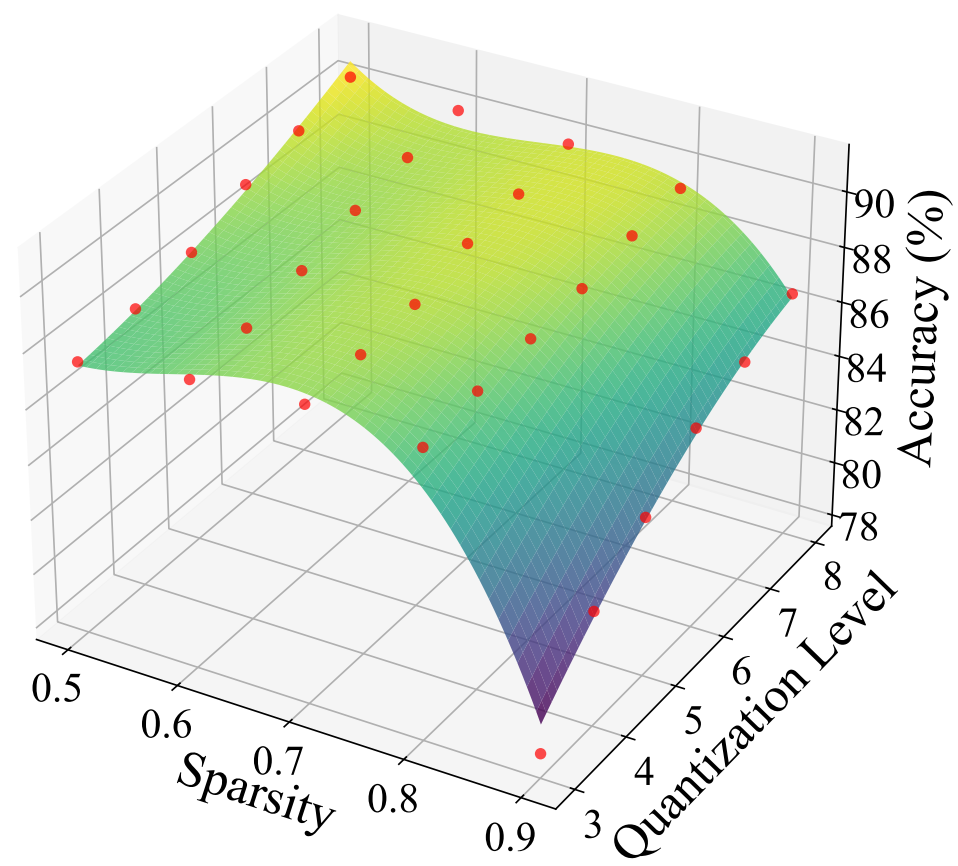}
\caption{Fitting results for SFT data processing accuracy.}
\label{FitPlot}
\end{figure}

The relationship between the objective accuracy function and the optimization parameters (sparsity rate and quantization level) is challenging to define theoretically due to the complex nature of model compression methods, such as pruning, quantization, and distillation. These methods modify the model's parameters and structure, leading to performance changes that are highly nonlinear, with intricate dependencies across network layers and the inherent non-linearity of activation functions.
To address this challenge, we approached the problem from a data-driven perspective. Specifically, we investigated how the accuracy is influenced by the sparsity rate and quantization level by fitting the observed data to a third-order function~\cite{Xinyu}. Through this approach, we derived a model that captures the relationship between accuracy and the optimization parameters. The fitting results, as illustrated in Fig. \ref{FitPlot}, demonstrate a strong alignment between the predicted and actual data, achieving a mean squared error of less than 0.26\%.

\section{Performance Analysis}\label{5.5}
\subsection{Fine-Tuning Delay Analysis}
In this section, we analyze the delay in each fine-tuning round, which includes several phases.

1)\ \emph{Transformer block distribution (TD) latency}: At the beginning of training, the server divides the transformer block into a server part and a device part. The device-side transformer block or aggregated device-side LoRA is broadcast to all devices. Let the initial data size of the allocated transformer block be denoted as $\Psi(l)$ and the data size of its corresponding LoRA as $\Psi^L(l)$.
Following Shannon's theorem, the rate of transformer block distribution is given by
$r = C \log_2\left( 1 + \mathrm{SNR}_s \right)$
where $C$ and $\mathrm{SNR}_s$ represent the total bandwidth and the signal-to-noise ratio, respectively. The corresponding transmission latency can be expressed as
\begin{equation}
\tau_{ED}^t(l) =
\begin{cases} 
{\Psi(l)}/{r}, & \text{if } t = 1; \\
{\Psi^L(l)}/{r}, & \text{if } t > 1.
\end{cases}
\end{equation}

2)\ \emph{Device-side local computation (CC) delay}:
The device's local computation delay involves the transformer block's FP delay, which includes both the pre-trained model and LoRA.
Let $\Phi^F_c (l)$ be the FLOPs required to be computed on device $n$, which will be discussed in the following subsection. 
The local computation delay for device $n$ can be given by
\begin{equation}
    \tau_{CC}^t (l) = { \Phi^F_c (l) }/{f_{n}C_n^uD_n^u},~\forall n \in \mathcal{N},
\end{equation}
where $f_n$ is the GPU frequency of device $n$, $C_n^U$ is the number of cores of the GPU at device $n$, and $D_n^U$  represents the number of FLOPs executed in a single core cycle of the GPU.

3)\ \emph{Immediate activation transmission (IT) latency}:
Each device transmits the immediate activation from the $l$-th transformer block to the server. Let $\Psi^A$ denote the size of the output, which matches the size of the input token. The transmission rate for device $n$ is given by $r_n^{UL}(b_n)= b_n \log_2\left( 1 + \mathrm{SNR}_n \right), \forall n \in \mathcal{N}$, where $b_n$ represent the allocated bandwidth to device $n$ and $\mathrm{SNR}_n$ represents its uplink signal-to-noise ratio. With the sparsification ratio $\beta$, the IT latency is defined as
\begin{equation}
   \tau_{\text{IA}}^t(\beta,b_n) = {\beta\Psi^A}/{r_n^{UL}(b_n)},~\forall n \in \mathcal{N}. 
\end{equation}

4)\ \emph{Server-side computation (SC) delay}:
The server's computation delay involves considering the forward and BP delay of the server-side transformer block. FP uses the pre-trained model along with the corresponding LoRA, while BP only updates the LoRA parameters, without updating the pre-trained model's parameters. Let $\Phi^F_s(l)$ be the total Flops needed to be computed in the FP of server and $\Phi^B_s(l)$ represent the computational load during LoRA's BP.
The training delay on server's training can be expressed as
\begin{equation}
     \tau_{SC}^t (l,f_n^s) = { (\Phi^F_s(l)+\Phi^B_s(l)) }/{f_n^sC_sD_s},~\forall n \in \mathcal{N}.
\end{equation}

\begin{figure}[!t]
\centering
\includegraphics[width=2.8in]{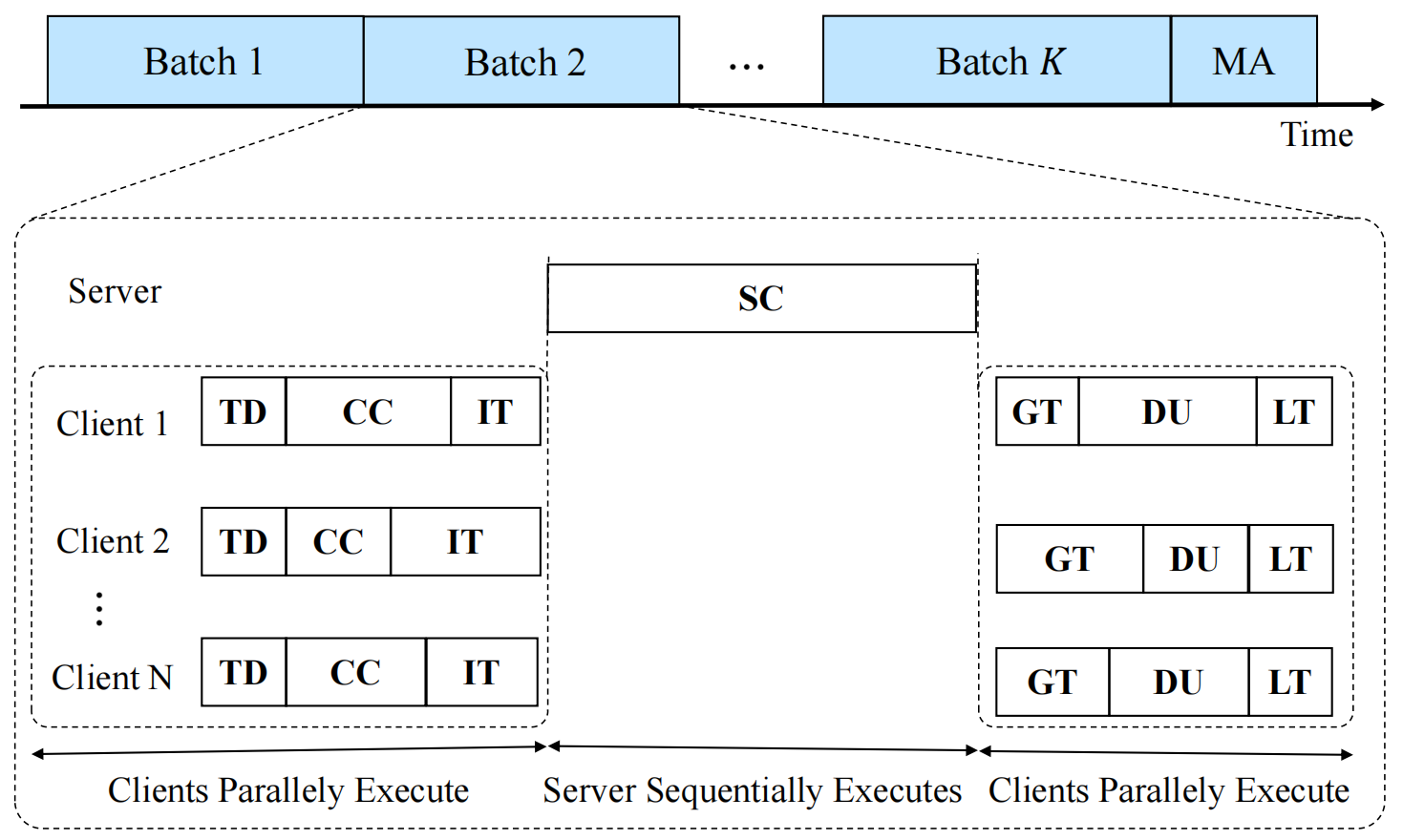}
\caption{Fine-tuning delay in each round.}
\label{Delay analysis}
\end{figure}

5)\ \emph{Gradient transmission (GT) latency}: The server transmits the gradient of immediate activation. We assume the transmission rate for the server is  $r_n^{DL}(b_n)= b_n \log_2\left( 1 + \mathrm{SNR}_s \right), \forall n \in \mathcal{N}$.
The gradient transmission delay from the server to the device can be given by 
\begin{equation}
   \tau_{\text{GT}}^t (\beta,b_n)= {\beta\Psi^G}/{r_n^{DL}(b_n)},~\forall n \in \mathcal{N}.  
\end{equation}

6)\ \emph{Device-side local updating (DU) delay }: 
Let $\gamma_c$ represent the computational load during LoRA's BP. The BP time of the server is given by
\begin{equation}
\tau_{\text{CU}}^t (l) = { \Phi^B_c (l) }/{f_{n}C_n^uD_n^u},~\forall n \in \mathcal{N}. 
\end{equation}

7)\ \emph{LoRA transmission (LT) latency}: Let $\Psi^L$ denote the size of device-side LoRA. The latency of uploading LoRA from device to the server can be represented as 
\begin{equation}
\tau_{\text{LT}}^t(b_n) = {\Psi^L(l)}/{r_n^{UL}(b_n)},~\forall n \in \mathcal{N}.
\end{equation}

Each round of the total training delay consists of multiple phases, including transformer block distribution delay, device-side local computation delay, immediate result transmission delay, server-side computation delay, gradient transmission delay, and device-side local updating delay. As shown in Fig.~\ref{Delay analysis}, the overall delay for each fine-tuning round can be expressed as
\begin{equation}
\begin{aligned}
    \tau_n^t(\beta,l,b_n,f_n^s) = &\tau_{\text{ED}} + \tau_{\text{CC}}(l) + \tau_{\text{IT}}(\beta,b_n) + \tau_{\text{SC}} (l)\\ &+ \tau_{\text{GT}} (\beta,b_n)+ \tau_{\text{DU}} (l,f_n)+\tau_{\text{LT}}(l,b_n).
\end{aligned}
\end{equation}
We assume that the set of bandwidths allocated to all devices is denoted as $\mathbf{b}$. As illustrated in the figure, the delay for each training round can be expressed as
\begin{equation}
\begin{aligned}
    \tau_t(\beta,l,\mathbf{b}) =\max\{\tau_n^t(\beta,l,b_n)\}, ~n\in\mathcal{N},
\end{aligned}
\end{equation}
and the total training delay after $T$ rounds is given by
\begin{equation}
\begin{aligned}
    \tau(\beta,l,\mathbf{b})=\sum_{t=1}^{T}\tau_t(\beta,l,\mathbf{b}).
\end{aligned}
\end{equation}

\subsection{Memory Consumption Analysis}
The memory consumption in each transformer block can be divided into four main components: model parameters, optimizer state, gradients, and activations. Each component's memory consumption depends on the number of parameters and the data type precision, which is represented by a variable \(\alpha\) (e.g., \(\alpha = 4\) for FP32 precision and \(\alpha = 2\) for FP16 precision).

\subsubsection{Model memory}
The memory required to store model parameters is determined by the parameter number and the precision. The total memory consumption can be expressed~as
\begin{equation}
M_{m} = \alpha (12D^2 + 18Dr).
\end{equation}

\subsubsection{Optimizer state}
The optimizer states require additional memory to store updating-related variables such as momentum and variance terms for each parameter, as well as optional parameter copies depending on the optimizer type. For example:
In the adaptive moment estimation (Adam) optimizer, the memory includes momentum and variance, requiring \(\hat{\alpha}  = 2\alpha\) bytes. If mixed-precision training is used, an additional parameter copy is needed, increasing the memory to \(\hat{\alpha}  = 3\alpha\).
In the stochastic gradient descent (SGD) optimizer, the memory only includes momentum, requiring \(\hat{\alpha}  = \alpha\).
The general memory requirement for optimizer states is therefore expressed as
\begin{equation}
M_{o} = \hat{\alpha}  (12D^2 + 18Dr).
\end{equation}

\subsubsection{Gradient memory}
The memory required for gradient storage is proportional to the parameter number and depends on the precision of the gradients. This can be expressed as:
\begin{equation}
M_{g} = \alpha (12D^2 + 18Dr).
\end{equation}

\subsubsection{Activation memory}
Activation memory depends on factors such as model size, recomputation strategies, and parallelization methods.  activation memory can be approximated~as
\begin{equation}
M_{a} = \alpha_1 BND + \alpha_2 BN^2A,
\end{equation}
where \(\alpha\) and \(\alpha_2\) are coefficients that depend on the specific model architecture and training setup when using the estimation formula from the Megatron scheme \(\alpha_1=34\) and \(\alpha_2=5\)~\cite{zhaojun}.

The total memory consumption in each transformer block can be expressed as 
\begin{equation}
\begin{aligned}
M_t &= M_m + M_o+M_g+M_a\\
&=(2\alpha+\hat{\alpha})(12D^2+18Dr)+ \alpha_1BND+\alpha_2BN^2A.
\end{aligned}
\end{equation}
The embedding layer's memory consumption for model weights and gradients is $4ND$, the output activations occupy $4B(N+1)D$, and there is additional memory consumption for local data of $4P^2CD$. The out layer's memory consumption is $4BND$. The memory consumption on the device side is expressed as
\begin{equation}
    M^c(l)=16D^2+BND+lM_t.
\end{equation}
\subsection{Communication Overhead and Computation Workload}
We first analyze the parameter size of the SFT scheme.
The number of parameters in each transformer block mainly comes from three parts: the multi-head self-attention (MSA), feed-forward network (FFN), and layer normalization (LayerNorm). In the MSA, weight matrices are independently created for the query, key, and value, and with the additional LoRA matrix parameters, the total number of parameters is $4   (D   r + r   D) + 4  D^2$. 
The FFN consists of two fully connected layers, where the first expands the embedding dimension \( D \) to \( D_{\text{mlp}} \) and the second reduces it back to \( D \). With LoRA, the total parameter number is \( 2(D  r + r  D_{\text{mlp}}) + 2D  D_{\text{mlp}} \), and given that \( D_{\text{mlp}} = 4D \), the parameter number simplifies to \( 8D^2 + 10D  r \).
Thus, the total number of parameters in each transformer block is approximately  $12D^2+18Dr$.
Additionally,
the number of parameters in the embedding layer is $(P^2C + N + 3)D$, which is related to the number and size of the patches. In the considered scheme, patches pass through multiple transformer blocks and are ultimately classified by a classification (CLS) output layer, which is typically a fully connected layer mapping an input of dimension \( D \) to \( K \) classes, resulting in \( D  K + K \) parameters.

\textbf{Computation workload}:
The computational FLOPs of the device and server are related to the deployed model and model parameters. In the proposed frame, an embedding layer and 
$l$ transformer blocks are deployed on the device side, while the remaining $(L-l)$ transformer blocks and the classification layer are deployed on the server side.
Each parameter typically requires 2 to 4 FLOPs in floating-point operations (one multiplication and one addition), and for simplicity, we approximate this as 2 FLOPs per parameter. 
Then, the FLOPs required for the device's FP are given by 
$
\Phi^F_c(l) = l(24BND^2+4BN^2D)+2BNDK
$
and the FLOPs required for BP are 
$
\Phi^B_c(l) =l(48BND^2+8BN^2D)+4BNDK.
$
The FLOPs required for the server's local forward and back propagation computations are 
$
\Phi^F_s(l)  = (L-l)(24BND^2+4BN^2D)
$
and
$
\Phi^B_s(l)= (L-l)(48BND^2+8BN^2D)+4BNDK,
$ respectively.

\textbf{Communication overhead}:
We further analyze the data transmission requirements for the ED, SC, and GT stages,  which will be discussed in the next subsection. During the ED stage, the pre-trained model and the initial LoRA model weights need to be distributed. Assuming each parameter occupies \( b \) bytes, the total data size to be transmitted in the ED stage is \(\Psi(l)= B  l (18Dr + 12D^2 + (P^2C + N + 3)D) \). In the SC stage, the intermediate patch sequence needs to be transmitted between the transformer blocks on the device and server sides, with a data size of \(BND \). In the GT stage, the size of the gradients to be transmitted is equal to the size of the intermediate activations.  During the LT stage, the size of the LoRA parameters to be transmitted is given by $\Psi^L(l)=2lBDr$.

\section{Problem Formulation}\label{5}
\subsection{Problem Formulation}
In the proposed optimization problem, the objective is to minimize the training delay while ensuring that the accuracy and memory constraints on devices are satisfied.  To achieve this, we focus on jointly optimizing LLM fine-tuning parameters, including the compression ratio and the number of transformer blocks executed locally, as well as efficiently allocating spectrum bandwidth resources among devices.
The final compression ratio \(\beta\) is jointly determined by sparsification, quantization, and encoding. In this optimization problem, we consider the tunable parameters \(\rho\) and \(E\) as the optimization variables.
The joint optimization problem can be expressed as follows. 
\begin{subequations} \label{eq:optimization_problem}
\begin{align}
    & \mathbb{P}: && \min_{\rho, E, l, \mathbf{b}} \quad \tau(\rho, E, l, \mathbf{b}) \label{eq:obj_func} \\
    & \text{s.t.} && \mathrm{A}(\rho, E) \geq \mathrm{A_{th}}, \label{eq:constraint_a} \\
    &              && \mathrm{M}^c(l) < M^c_\mathrm{max},\quad \forall l \in \mathcal{L}, \label{eq:constraint_b} \\
    &              && \rho_\mathrm{min} \leq \rho \leq \rho_\mathrm{max} , \label{eq:constraint_rho} \\
    &              && E_\mathrm{min} \leq E \leq E_\mathrm{max},, \label{eq:constraint_E} \\
    &              && b_n \leq b_n^\mathrm{max},\quad \forall n \in \mathcal{N}, \label{eq:constraint_bn} \\
    &              && \sum _{n \in \mathcal{N}} b_n = b_s^\mathrm{max}. \label{eq:constraint_sum_bn}
\end{align}
\end{subequations}
Constraint~\eqref{eq:constraint_a} ensures that the system's accuracy meets the minimum requirement, where \( A_\mathrm{th} \) represents the maximum accuracy achievable by the LLM within an allowable accuracy degradation. Constraint~\eqref{eq:constraint_b} enforces the storage resource usage on devices to remain within allowable limits. Constraints~\eqref{eq:constraint_rho} and~\eqref{eq:constraint_E} bound the sparsity rate \(\rho\) and quantization level \(E\) within their respective ranges. Finally, constraints~\eqref{eq:constraint_bn} and~\eqref{eq:constraint_sum_bn} ensure that both individual bandwidth allocations and the total bandwidth allocation comply with system-wide limitations, maintaining resource balance and operational feasibility.

\subsection{Problem Decomposition}
To efficiently solve the joint optimization problem presented in Eq. \eqref{eq:optimization_problem}, we decompose it into two subproblems by leveraging the hierarchical structure of the decision variables. The first subproblem focuses on optimizing the fine-tuning configuration, including the cut layer \(l\), sparsification rate \(\rho\), and quantization level \(E\), to minimize the overall fine-tuning delay while satisfying accuracy and resource constraints. The second subproblem focuses on optimizing the network bandwidth allocations to minimize the fine-tuning delay for a given fine-tuning configuration.

\subsubsection{Subproblem 1 - Joint optimization of sparsification rate, quantization level, and allocated block}

The first subproblem addresses the optimization of the cut layer \(l\), sparsification rate \(\rho\), and quantization level \(E\), which directly impact the training process. The goal is to minimize the overall training delay while ensuring the system accuracy meets the specified requirements and resource constraints are satisfied. The subproblem can be formulated as follows:
\begin{subequations} \label{eq:subproblem_1}
\begin{align}
    & \mathbb{P}_1: && \min_{\rho, E, l} \quad \tau(\rho, E, l) \\
    & \text{s.t.} && \mathrm{A}(\rho, E) \geq \mathrm{A_{th}}, \\
    &              && \mathrm{M}^c_n(l) < M^c_\mathrm{max}, \quad \forall  n \in \mathcal{N}, \\
    &              && \rho_\mathrm{min} \leq \rho \leq \rho_\mathrm{max}, \\
    &              && E_\mathrm{min} \leq E \leq E_\mathrm{max}, \\
    &              && 0 < l < L.
\end{align}
\end{subequations}

\subsubsection{Subproblem 2 - Optimization of spectrum bandwidth allocation}
The second subproblem focuses on the dynamic bandwidth allocation \(\mathbf{b}\) of network resources. These parameters are optimized based on the training configuration (\(l^\star\), \(\rho^\star\), and \(E^\star\)) obtained from Subproblem \(\mathbb{P}_1\). The objective is to minimize the training delay by effectively allocating network resources in response to the dynamic environment. The subproblem can be formulated as follows:
\begin{subequations} \label{eq:subproblem_3}
\begin{align}
    & \mathbb{P}_2: && \min_{\mathbf{b}} \max_{n \in \mathcal{N}} \quad \tau_t(\mathbf{b_n}) \\
    & \text{s.t.} && b_n \leq b_n^\mathrm{max}, \quad \forall n \in \mathcal{N}, \\
    &              && \sum_{n \in \mathcal{N}} b_n = b_s^\mathrm{max}.
\end{align}
\end{subequations}

\begin{algorithm}[t]
\caption{Lagrange-based resource management algorithm.}
\label{alg:solution_P1}

\KwIn{Discrete variable range \( l \in \mathcal{L} \), initial Lagrange multipliers \( \lambda_0 \), step size \( \mu_k \), tolerance \( \epsilon \)}
\KwOut{Optimal solution \( (\rho^\star, E^\star, l^\star) \)}

Initialize Lagrange multipliers \( \lambda = \lambda_0 \)\;
Set iteration counter \( k = 0 \)\;

\Repeat{convergence}{
    \For{each \( l \in \mathcal{L} \)}{
        Fix \( l \) and optimize the continuous variables \( \rho \) and \( E \) by maximizing \( \mathcal{L}(\rho, E, l, \lambda) \)\;
        Obtain the optimal \( (\rho_l^\star, E_l^\star) \) for the fixed \( l \)\;
        Compute \( \mathcal{L}(\rho_l^\star, E_l^\star, l, \lambda) \) and store \( \{\rho_l^\star, E_l^\star, \mathcal{L}(\rho_l^\star, E_l^\star, l, \lambda)\} \)\;
    }
    Select \( l^\star = \arg \max_{l} \mathcal{L}(\rho_l^\star, E_l^\star, l, \lambda) \) and corresponding \( (\rho^\star, E^\star) = (\rho_{l^\star}^\star, E_{l^\star}^\star) \)\;

    Update the Lagrange multipliers:
    \[
    \lambda_i^{k+1} = \lambda_i^k + \mu_k \, g_i(\rho^\star, E^\star, l^\star)
    \]
    where \( g_i(\rho^\star, E^\star, l^\star) \) represents the degree of constraint violation\;

  \If{\( |\mathcal{L}(\rho^\star, E^\star, l^\star, \lambda) - \mathcal{L}_{\text{prev}}| < \epsilon \) \textbf{and} all constraints are satisfied}{
        Convergence achieved\;
        \textbf{break}\;
    }
    Update \( k = k + 1 \)\;
}

\end{algorithm}

\subsubsection{Relationship between subproblems}

The two subproblems are decomposed based on large and small timescales and solved sequentially. First, in the large timescale, \(\mathbb{P}_1\) is solved to determine the optimal training configuration, including \(l^\star\), \(\rho^\star\), and \(E^\star\), as the model structure remains consistent across training rounds. These results are then used as fixed inputs in the small timescale to solve \(\mathbb{P}_2\), where network resource allocation is optimized based on device-specific heterogeneous resources and dynamic channel conditions.
This decomposition and sequential optimization effectively reduce computational complexity by clearly separating global and real-time optimization tasks, enabling efficient algorithm design tailored to system accuracy, resource utilization, and dynamic performance requirements.

\begin{algorithm}[t]
\caption{SQP-based spectrum bandwidth allocation algorithm.}
\label{alg:sqp_algorithm}

\KwIn{Initial feasible solution \((\mathbf{b}^0, \tau^{\star,0})\), maximum iterations \(K_{\text{max}}\), tolerance \(\epsilon\)}
\KwOut{Optimized solution \((\mathbf{b}^\star, \tau^\star)\)}

Initialize \(k = 0\)\;

\Repeat{convergence}{
    Compute the objective value \(\tau_n^k = \tau(b_n^k)\) for all \(n \in \mathcal{N}\)\;

    Compute gradients \(\nabla_{\mathbf{b}} \tau_n^k\) at the current iterate\;

    Formulate the QP subproblem by linearizing the constraints as in Eq.~\eqref{eq:qp_subproblem}\;

    Solve the QP subproblem to obtain \(\Delta \mathbf{b}\) and \(\Delta \tau^\star\)\;

    Update variables:
    \[
    \mathbf{b}^{k+1} = \mathbf{b}^k + \Delta \mathbf{b},
    \tau^{\star,k+1} = \tau^{\star,k} + \Delta \tau^\star
    \]

    Check convergence criteria: \If{\(|\tau^{\star,k+1} - \tau^{\star,k}| \leq \epsilon\) \textbf{or} \(\|\Delta \mathbf{b}\|_2 \leq \epsilon\)}{
        Convergence achieved\;
        \textbf{break}\;
    }

    Update \(k = k + 1\)\;

    \If{\(k \geq K_{\text{max}}\)}{
        \textbf{break}\;
    }
}

\end{algorithm}

\section{Proposed Solution}\label{6}

\subsection{Fine-Tuning Configuration Optimization}
To simplify the subproblem $\mathbb{P}_1$, we introduce Lagrangian multipliers \( \lambda \) to relax some of the constraints, forming the Lagrangian function \( \mathcal{L}(\rho, E, l, \lambda) \) as follows:
\begin{equation}
\begin{aligned}
     \mathcal{L}(\rho, E, l, \mathbf{\lambda}) &= \tau(\rho, E, l) + \lambda_1 (\mathrm{A}(\rho, E) - \mathrm{A_{th}}) + \\
     &\lambda_2 (\mathrm{M}^c_\mathrm{max} - \mathrm{M}^c(l)),
\end{aligned}
\end{equation}
where \( \mathbf{\lambda} = [\lambda_1, \lambda_2] \) are Lagrangian multipliers that penalize the violation of accuracy and resource constraints, with \( \lambda_1, \lambda_2 \geq 0 \). By relaxing these constraints, we convert the original problem into an unconstrained optimization problem and aim to maximize the relaxed Lagrangian function.

Since \( l \) is a discrete variable with a finite range, we can decompose the problem into subproblems for different values of \( l \). For each possible value of \( l \), we optimize the continuous variables \(\rho\) and \(E\), and select the combination that maximizes the objective function. The process can be expressed as follows:

\subsubsection{Solving the discrete variable via exhaustive search}
For each \( l \in \{1, 2, \dots, L\} \), we can use methods like exhaustive search to evaluate the optimal solution for each discrete value of \( l \). The solution process begins by fixing \( l \), which involves selecting a specific value for \( l \). With \( l \) fixed, the next step is to optimize the continuous variables \(\rho\) and \(E\) by maximizing the Lagrangian function \( \mathcal{L}(\rho, E, l, \lambda) \). After determining the optimal \(\rho\) and \(E\) for the fixed \( l \), we compute and record the corresponding values of \( l \), \(\rho\), and \(E\) along with the Lagrangian function value. Thus, we can identify the optimal \(\rho\), \(E\), and the maximum objective function value corresponding to each possible value of \( l \).

\subsubsection{Optimizing the continuous variables}
For a fixed value of \( l \), optimizing the continuous variables \(\rho\) and \(E\) reduces to a single-variable optimization problem for each. This can be solved using several methods. If the Lagrangian function \( \mathcal{L}(\rho, E, l, \lambda) \) is differentiable with respect to \(\rho\) and \(E\), we can apply gradient ascent to iteratively update \(\rho\) and \(E\) in the direction that increases \( \mathcal{L}(\rho, E, l, \lambda) \), thereby maximizing the function. Another method is to use convex optimization techniques, which are particularly efficient if \( \mathcal{L}(\rho, E, l, \lambda) \) is a convex function in \(\rho\) and \(E\). By using convex optimization, we can quickly and reliably solve for the optimal \(\rho\) and \(E\) when convexity is present. Additionally, grid search can be used when the function's differentiability or convexity is not guaranteed.

\subsubsection{Updating the Lagrange multipliers}
To ensure that the relaxed solution satisfies the original problem's constraints, we use the subgradient method to iteratively update the Lagrange multipliers. We denote all the inequality constraints by $g_p(\rho,E)=\mathrm{A}(\rho,E)-\mathrm{A_{th}} \geq 0$ for $q=1$, $g_p(l)= M^c_\mathrm{max} -\mathrm{M}^c_n(l)\geq 0$ for $p=2,\dots,N+1$, $g_p(\rho)= \rho -\rho_\mathrm{min}\geq 0$ for $p=N+2$, $g_p(\rho)= \rho_\mathrm{max} -\rho\geq 0$ for $p=N+3$, $g_p(E)= E -E_\mathrm{min}\geq 0$ for $p=N+4$, $g_p(E)= E_\mathrm{max} -E\geq 0$ for $p=N+5$, $g_p(l)= L -l\geq 0$ for $p=N+6$, and $g_p(l)= l \geq 0$ for $p=N+7$. 
The update rule is given by
\[
\lambda_i^{k+1} = \lambda_i^k + \mu_k \, g_i(\rho, E, l),
\]
where \( \mu_k \) is the step size. During each iteration, the Lagrange multipliers are adjusted based on the extent of violation of each constraint, ensuring that the solution progressively satisfies the constraints as the process converges.

\subsection{Resource Allocation Optimization}

Problem $P_2$ optimizes the fine-tuning delay by dynamically allocating bandwidth $\mathbf{b}$. The objective function $\tau(\mathbf{b})$ in $P_2$ is nonlinear, and the constraints include both equality and inequality conditions. Due to the inherent nonlinearity of the objective function and the complexity of the constraints, we adopt the SQP method to solve this problem. The process involves reformulating the problem to eliminate the nested max operator, followed by constructing and solving Quadratic Programming (QP) subproblems at each iteration.

\subsubsection{Reformulation of the problem}
The original objective function \(\min \max \tau(b_n)\) involves a nested \(\max\) operator that complicates direct optimization. To simplify, we introduce an auxiliary variable \(t\) and replace the \(\max\) term with an inequality constraint:
\begin{equation}
    \tau^\star \geq \tau(b_n), \quad \forall n \in \mathcal{N}.
\end{equation}
The optimization problem is reformulated as
\begin{subequations} \label{eq:subproblem_2}
\begin{align}
    & \mathbb{P}_3: && \min_{\mathbf{b}}~~~\tau^\star\\
    & \text{s.t.} && b_n \leq b_n^\mathrm{max}, \quad \forall n \in \mathcal{N}, \\
    &              && \sum_{n \in \mathcal{N}} b_n = b_s^\mathrm{max}\\
    &              && \tau^\star \geq \tau(b_n), \quad \forall n \in \mathcal{N}.
\end{align}
\end{subequations}
The reformulated problem \(\mathbb{P}_3\) minimizes the auxiliary variable \(\tau^\star\) subject to linear and nonlinear constraints. The inequality constraint \(\tau^\star \geq \tau(b_n)\) is nonlinear and must be linearized at each iteration to construct the QP subproblem.

\subsubsection{Linearization of the nonlinear constraint}

To construct the QP subproblem, we need to linearize the nonlinear inequality constraint \(\tau^\star \geq \tau(b_n)\) around the current iterate \(\mathbf{b}^k\). Using a first-order Taylor expansion, the constraint can be approximated as:
\begin{equation}
    \tau^\star \geq \tau(b_n^k) + \nabla_{\mathbf{b}} \tau_n^k (\mathbf{b} - \mathbf{b}^k), \quad \forall n \in \mathcal{N},
\end{equation}
where \(\nabla_{\mathbf{b}} \tau_n^k\) is the gradient of \(\tau(b_n)\) with respect to \(\mathbf{b}\) evaluated at \(\mathbf{b}^k\).

\subsubsection{Construction of the QP subproblem}
At each iteration \(k\), the QP subproblem is formulated by approximating the original problem \(\mathbb{P}_3\) with a quadratic objective function and linear constraints:
\begin{subequations} \label{eq:qp_subproblem}
\begin{align}
    & \mathbb{P}_4: && \min_{\Delta \mathbf{b}, \Delta \tau^\star}~~~\tau^\star\\
    & \text{s.t.} && 0 \leq b_n^{k} + \Delta b_n \leq b_n^\mathrm{max}, \quad \forall n \in \mathcal{N}, \\
    &              &&  \sum_{n \in \mathcal{N}} (b_n^{k} + \Delta b_n) = b_s^\mathrm{max},\\
    &              && \tau^{\star,k} + \Delta \tau^\star \geq \tau_n^k + \nabla_{\mathbf{b}} \tau_n^k \Delta \mathbf{b}, \quad \forall n \in \mathcal{N}.
\end{align}
\end{subequations}
where \(\Delta \mathbf{b} = \mathbf{b} - \mathbf{b}^k\), and \(\Delta \tau^\star = \tau^\star - \tau^{\star,k}\).

\subsubsection{Solving the QP subproblem}
The QP subproblem defined in (\ref{eq:qp_subproblem}) is a convex optimization problem with linear constraints and a linear objective function, which can be efficiently solved using standard QP solvers like the interior-point method or active-set method. The steps to solve the QP subproblem are:

\begin{itemize}
    \item {Gradient computation}: At the current iteration \(k\), compute the gradient \(\nabla_{\mathbf{b}} \tau_n^k\) for each device \(n\), which is necessary for constructing the linearized constraints.

    \item {Formulation of the QP problem}: Use the computed gradient and the current iterate \((\mathbf{b}^k, \tau^{\star,k})\) to formulate the QP subproblem as described in (\ref{eq:qp_subproblem}). This involves substituting the linearized constraints into the objective function and ensuring all resource allocation constraints are satisfied.

    \item {Solve the QP problem}: Employ a standard QP solver (e.g., an interior-point method or active-set method) to solve the convex optimization problem. The solver provides the optimal increments \(\Delta \mathbf{b}\) and \(\Delta \tau^\star\).

    \item {Update variables}: Update the current variables using the obtained increments:
    $\mathbf{b}^{k+1} = \mathbf{b}^k + \Delta \mathbf{b},$
    and
    $\tau^{\star,k+1} = \tau^{\star,k} + \Delta \tau^\star.$

    \item {Check convergence}: Evaluate the convergence criteria. If the change in the objective function \(|\tau^{\star,k+1} - \tau^{\star,k}|\) is less than a predefined tolerance \(\epsilon\), or if the norm of the step sizes \(\|\Delta \mathbf{b}\|_2 + |\Delta \tau^\star|\) is sufficiently small, the algorithm terminates. Otherwise, proceed to the next iteration.

    \item {Repeat}: Iterate the process from Step 1 until the convergence criteria are met or the maximum number of iterations \(K_{\text{max}}\) is reached.
\end{itemize}

\begin{figure*}[t]
\centering
\subfloat[CIFAR100 IID]{\includegraphics[width=1.6in]{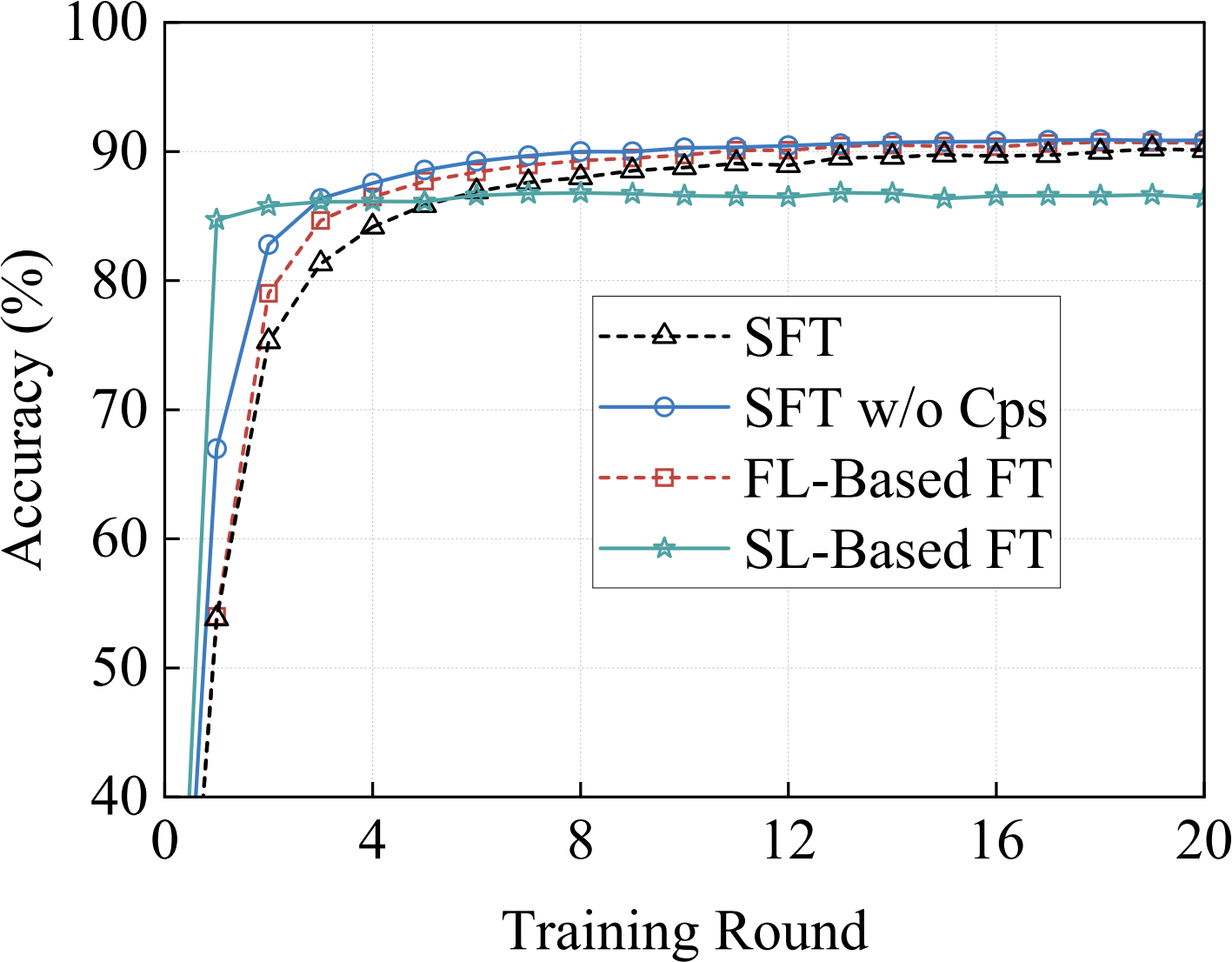}\label{cifar100_iid}}
\subfloat[CIFAR100 non-IID]{\includegraphics[width=1.6in]{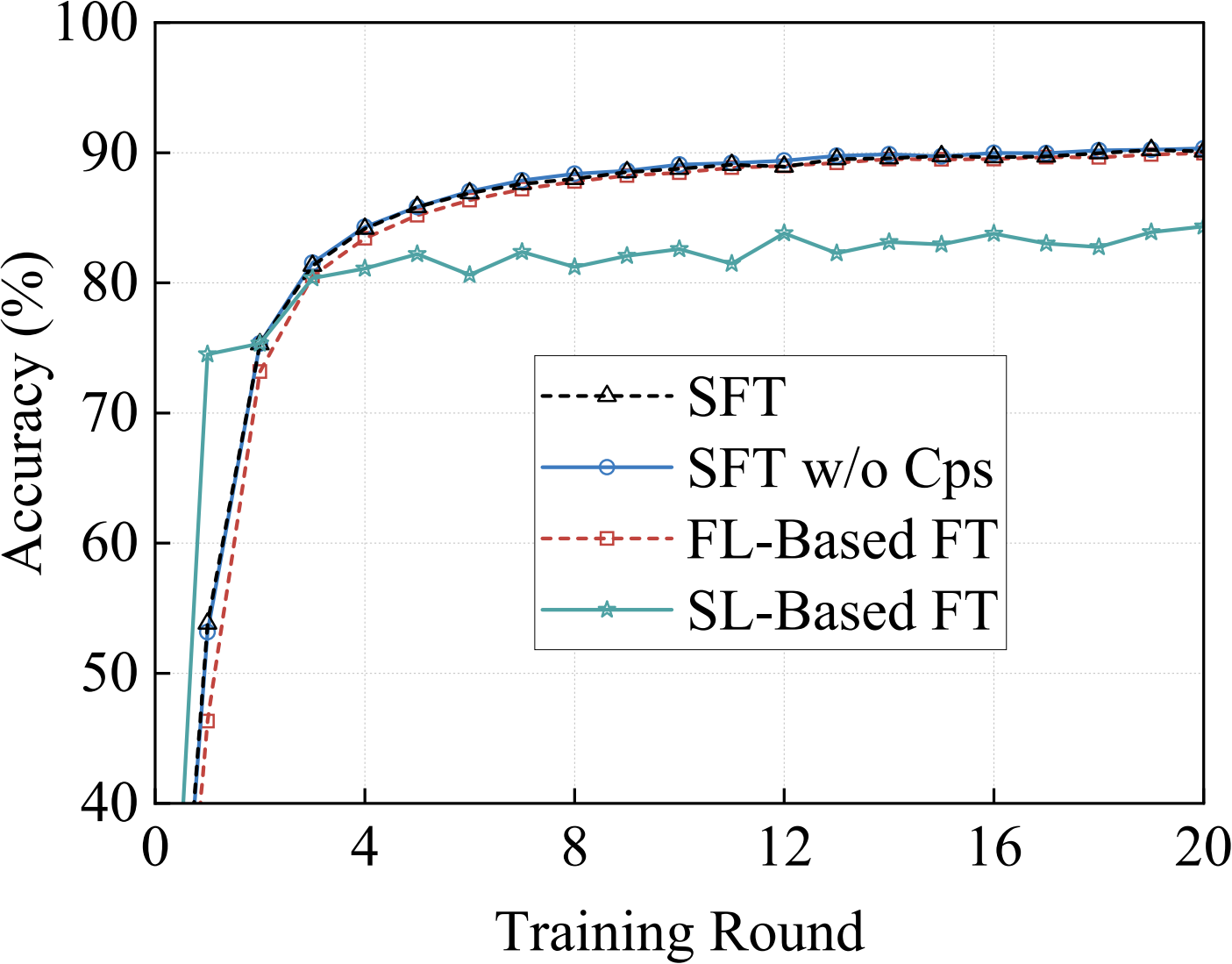}\label{cifar100_noniid}}
\subfloat[Tiny-ImageNet IID]{\includegraphics[width=1.6in]{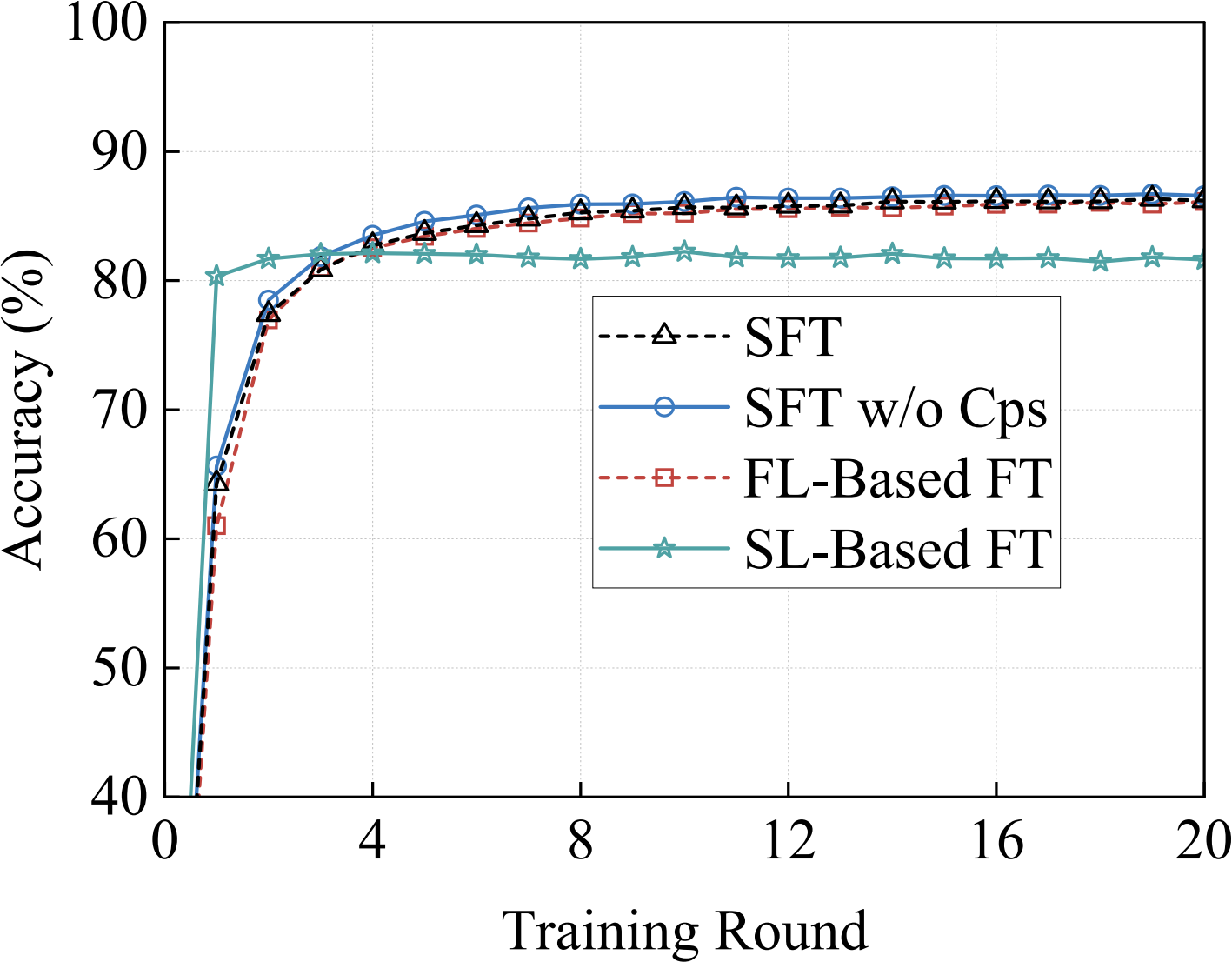}\label{tinyimagenet_iid}}
\subfloat[Tiny-ImageNet non-IID]{\includegraphics[width=1.6in]{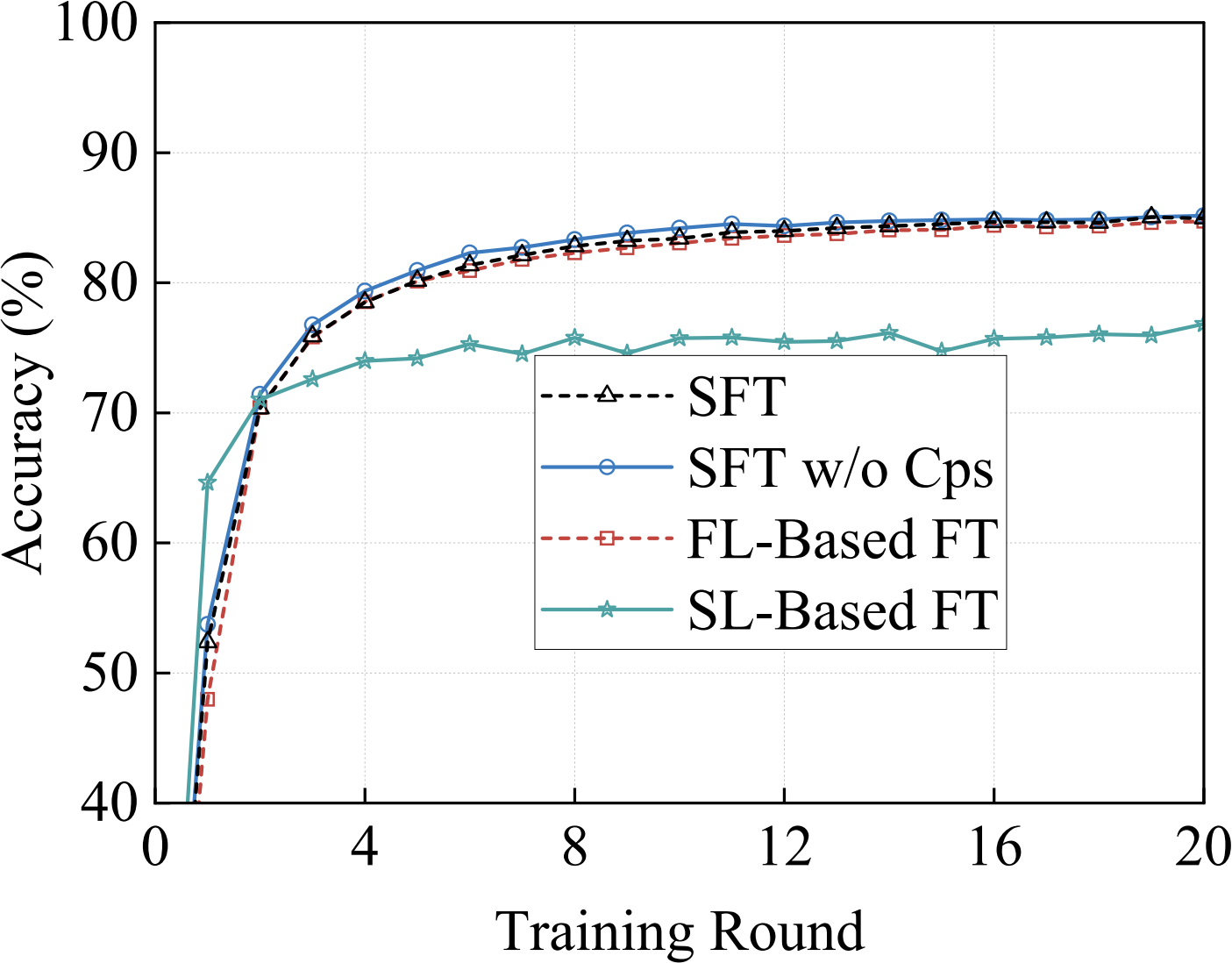}\label{tinyimagenet_noniid}}
\caption{Fine-tuning performance comparison among different schemes.}
\label{result1}
\end{figure*}

\begin{table}[t]
\centering
\caption{Simulation Parameter}
\label{tab:experiment_parameters}
\begin{tabular}{l|c|l|c}
\toprule
\textbf{Parameter}             & \textbf{Value}       & \textbf{Parameter}             & \textbf{Value}       \\ \midrule
$B$              & 30 MHz               & \# devices                        & 8                    \\ 
$D_n^u$               & 4                    & \# iteration Rounds             & 20                   \\ 
$C_n^u$       & $2 \times 10^9$      & LLM                          & ViT-base~\cite{VIT16}             \\ 
$C_s$            & 1.5 GHz              & Cut layer point                & 5-th ViT block       \\ 
$D_s$              & 2,048                 & \# model parameters             & 86 M                  \\ 
$f^s$       & $3 \times 10^9$      & Optimizer                      & SGD                  \\ 
Learning rate                  & 1e-4                 & Momentum                       & 0.9                  \\ 
Decay coefficient                     & 0.998                & Batch size                     & 64                   \\ 
Local epoch                    & 1                    & LoRA rank                      & 16                   \\ 
\bottomrule
\end{tabular}
\end{table}

\begin{table}[t]
\centering
\caption{Memory Comparison on Device Side among Different Schemes.}
\begin{tabular}{ccccc}
\toprule
\begin{tabular}[c]{@{}c@{}}Memory \\ (MB)\end{tabular} & \makecell{FL-Based \\ FT} & \makecell{FL-\\LoRA} & \makecell{SL-Based \\ FT} & Proposed \\ \midrule
Input data                                                       & 36.75       & 36.75   & 36.75       & 36.75 \\
Activation                                                       & 9171.11     & 9513.82 & 3941.23     & 3941.23 \\
Model+LoRA                                                       & 327.88      & 337.03  & 141.19      & 141.19 \\
Optimizer (SGD)                                                   & 329.76      & 9.09    & 3.75        & 3.75   \\ \midrule
Total                                                            & \makecell{9865.5\\(2.39$\times$)}      & \makecell{9896.69 \\(2.4$\times$)} & \makecell{4122.92\\(1$\times$)}     & \makecell{\textbf{4122.92}\\ (\textbf{1$\times$)}} \\ \bottomrule
\end{tabular}
\label{tab:memory_usage}
\end{table}

\section{Simulation Results}
\label{7}
\subsection{Experiment Setting}
For the experimental evaluations, we focus on the image classification task using the CIFAR100 dataset~\cite{krizhevsky2009learning}, which contains 50,000 training samples and 10,000 test samples across 100 classes, and the Tiny-ImageNet dataset~\cite{Tiny-ImageNet}, which includes 100,000 training samples and 10,000 validation samples spanning 200 classes. The details of the parameter and hyper-parameter settings for model training are presented in Table \ref{tab:experiment_parameters}.  For the independent and identically distributed (IID) setting, we randomly partition the dataset into multiple shards and uniformly assign them to each device. For the non-IID setting, we use the Dirichlet distribution with a concentration parameter of 0.5 for the dataset partition.

In the default experimental setting, the simulated SFT architecture consists of 1 edge server and 8 devices. We consider a system bandwidth of 5 MHz, with an SNR of 17 dB between the devices and the server. The computation frequency of the devices is randomly generated between 0.5 GHz and 1.5 GHz, while the server's computation frequency is set to 40 GHz. For the large model fine-tuning setup, the batch size is set to 64, the hidden layer dimension is configured to 3072, and the image size is set to \( 224 \times 224 \) with each patch size of \( 16 \times 16 \). We adopt the mobile devices are NVIDIA Jetson Nano equipped with a 256-core GPU and a floating-point operation capacity of 4. The server is equipped with advanced GPUs featuring 2048 cores and a floating-point operation capacity of 4.

We compare the proposed SFT with the distributed learning schemes as follows. To ensure fairness, we follow the same hyper-parameter setting for various baseline schemes.
\begin{itemize}
    \item {\bf FL-Based Fine-Tuning (FT)}: The traditional FL scheme proposed by Google in \cite{mcmahan2017communication}. By incorporating LoRA tuning into FL, each device transmits only low-rank matrices to the server, significantly reducing communication overhead.
    \item {\bf SL-Based FT}:
    SL divides a neural network into two parts, typically split between devices and a central server, thereby reducing training overhead by offloading computations and minimizing data transfer and local processing demands on the device side. \cite{SL7} 
    \item {\bf SFT w/o Compression}: The SFT  operates without the joint compression scheme.
\end{itemize}

\begin{figure}[t]
\centering
{\includegraphics[width=2.25in]{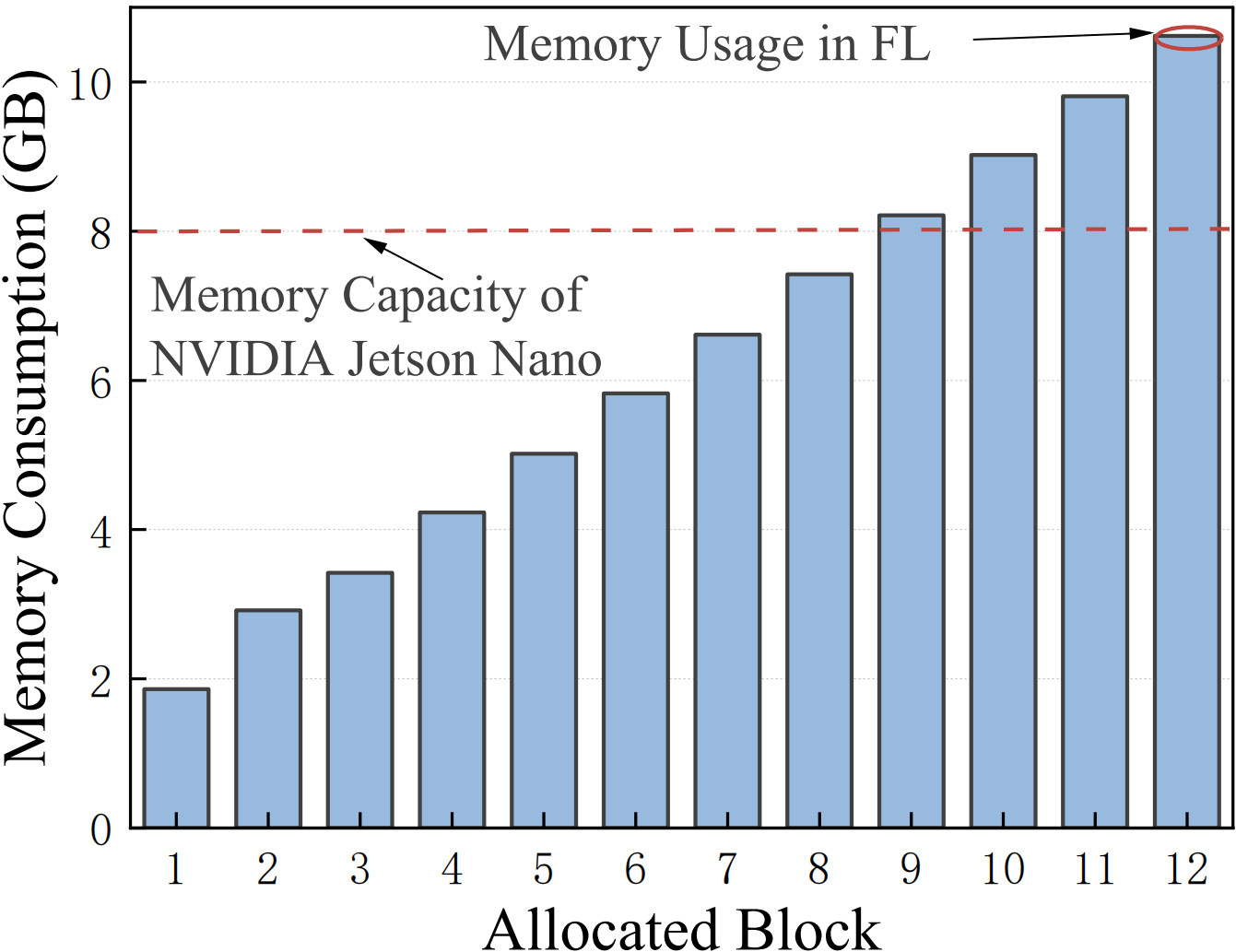}}
\caption{Memory consumption on a device with respect to allocated ViT block.}
\label{memory}
\end{figure}

\begin{figure}[t]
\centering
\includegraphics[width=2.2in]{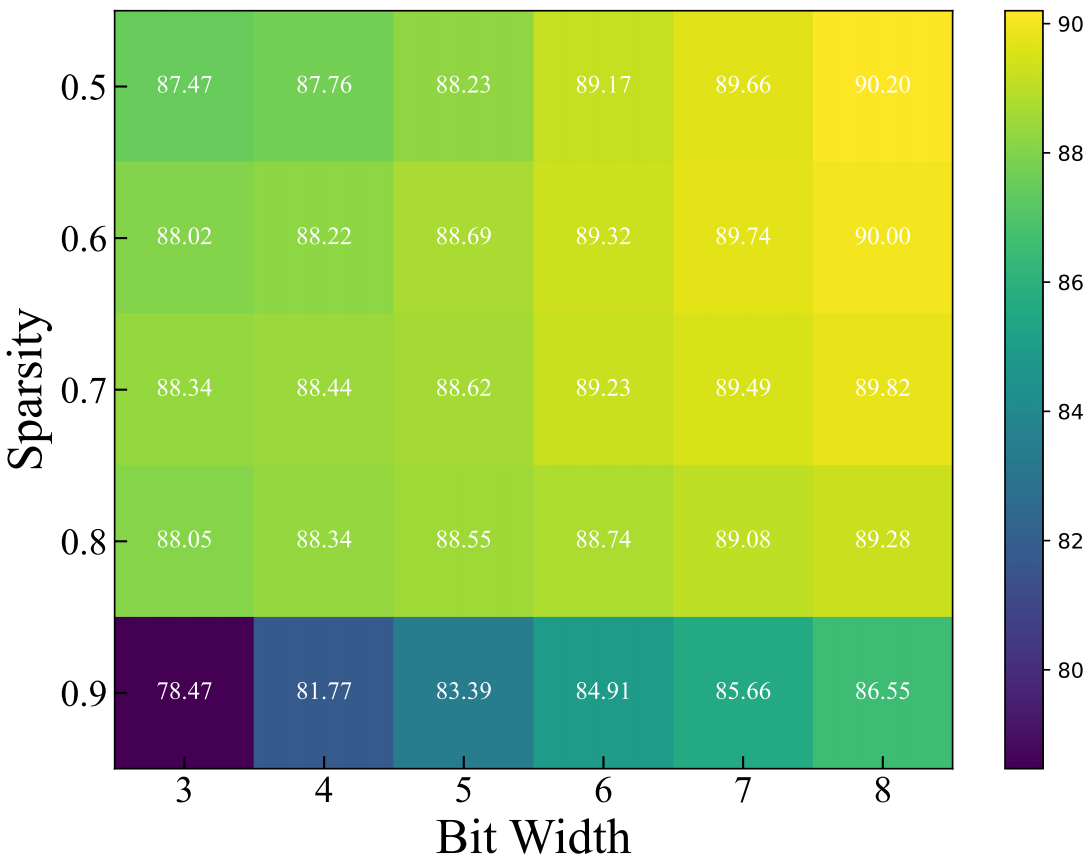}\label{compression1}
\caption{Impact of sparsity and bit width on accuracy.}
\label{CompressionResult}
\end{figure}

\begin{figure}[t]
\centering
\subfloat[Communication overhead in different fine-tuning schemes.]{\includegraphics[width=1.67in]{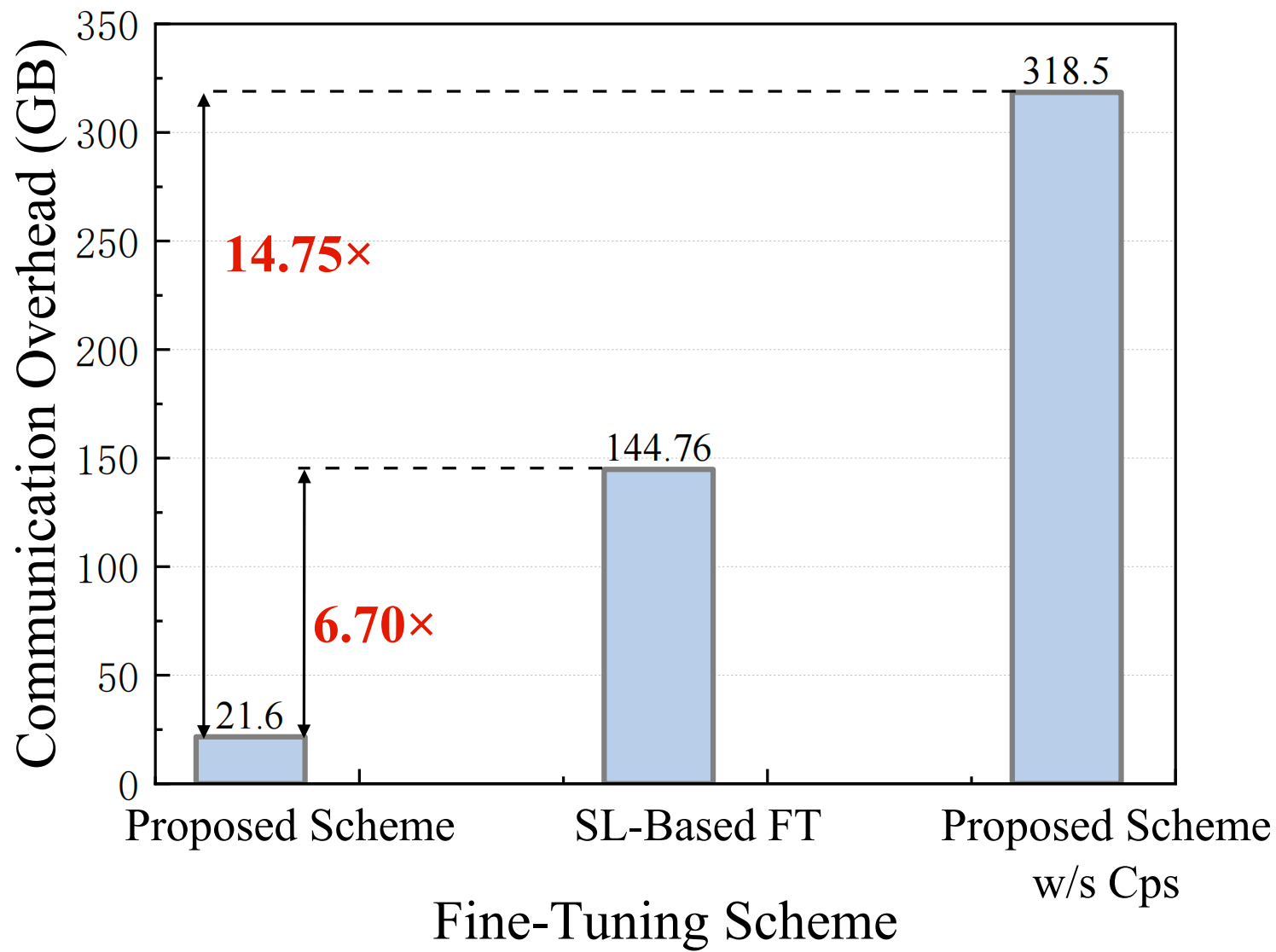}\label{compression4}}
\hfil
\subfloat[Per-round gains achieved by different compression schemes.]{\includegraphics[width=1.55in]{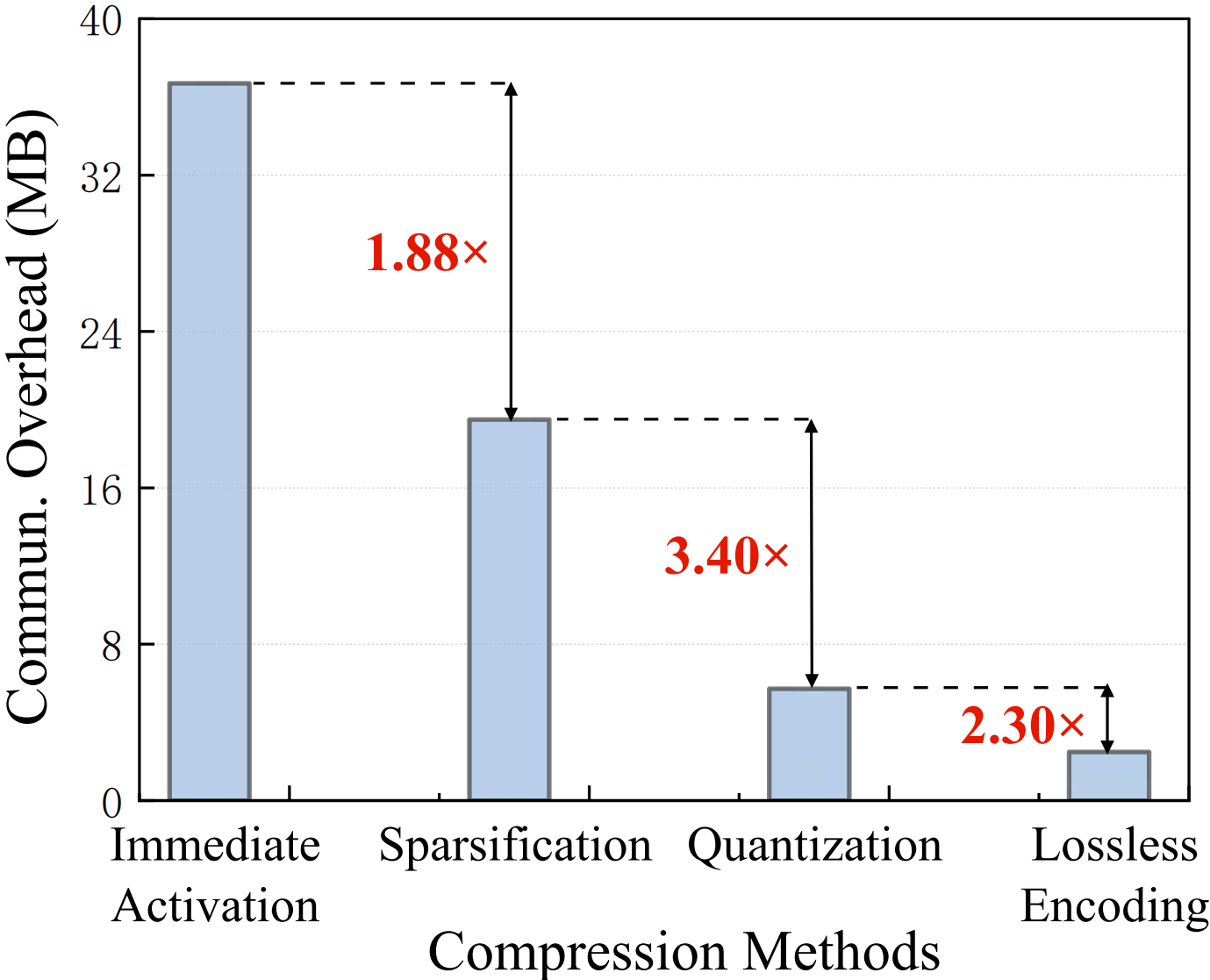}\label{compression3}}\\
\caption{Communication overhead in different fine-tuning schemes and different compression methods.}
\label{CompressionResult2}
\end{figure}

\begin{figure}[t]
\centering
\includegraphics[width=2.4in]{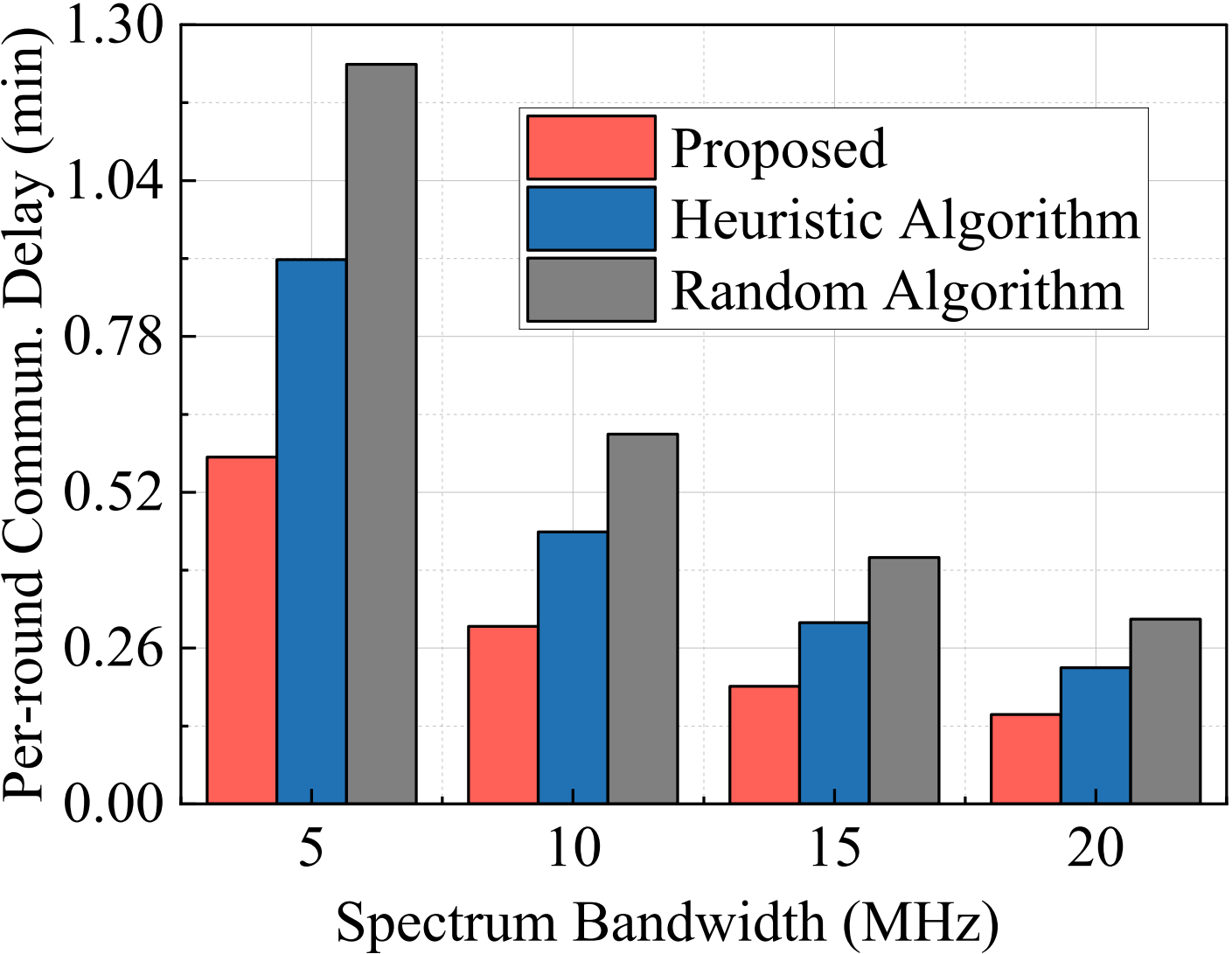}
\caption{Fine-tuning delay performance comparing among different optimization schemes.}
\label{Delay1}
\end{figure}

\subsection{Performance Evaluation}
\subsubsection{Fine-Tuning Accuracy} Fig. \ref{result1} shows the simulation results of the training process, highlighting several key observations. 
First, the proposed SFT scheme demonstrates robust convergence under both IID (in Fig. \ref{cifar100_iid} and Fig. \ref{tinyimagenet_iid}) and non-IID conditions (in Fig. \ref{cifar100_noniid} and Fig. \ref{tinyimagenet_noniid}).
Second, the accuracy achieved by our scheme is comparable to that of state-of-the-art benchmarks. 
Referring to the observation in the dropout layer, the distorted intermediate activations not only enhance the robustness of model training, but also lead to slight improvements in accuracy.
Similarly, compressing the intermediate activations is expected to achieve a comparable accuracy.
The results confirm that the efficiency improvements achieved through compression are not at the expense of training performance.

\subsubsection{Memory Consumption}
Table \ref{tab:memory_usage} presents the peak memory usage on the device side under different schemes. The results show that the proposed SFT and SL schemes reduce memory consumption by 58.2\% compared to the FL scheme, where 5 ViT blocks were allocated on the device side. 
Additionally, we evaluate the performance of LoRA applied to the FL scheme and find that LoRA does not effectively reduce memory consumption. This is because activations account for a significant portion of memory consumption, while the optimizer, which LoRA reduces, contributes only a small fraction to the total memory consumption. Therefore, when fine-tuning LLM on the device, it is essential to split the LLM to address memory constraints on devices.
Fig.~\ref{memory} illustrates the relationship between the required memory size on the device and the number of allocated ViT blocks, showing a proportional increase. The red line in the figure represents the maximum memory capacity of the NVIDIA Jetson Orin Nano. It is evident that the FL scheme exceeds this memory limit, making it unsuitable for deployment on resource-constrained devices.

\subsubsection{Communication Overhead} The performance of the proposed compression scheme is shown in Fig. \ref{CompressionResult}. The results demonstrate that the scheme can compress the transmission volume of intermediate activations within an allowable accuracy degradation. However, the compression cannot be arbitrarily small, as we observe that accuracy starts to drop sharply when the sparsity reaches 90\%. To limit the accuracy loss within 2\%, the sparsity and random quantization methods can achieve a maximum compression ratio of 12x with 80\% sparsity and 3-bit quantization, which can be further increased to 20x when combined with lossless encoding.

The impact of different compression schemes on communication overhead, as well as the communication overhead of various fine-tuning methods, is illustrated in Fig. \ref{CompressionResult2}. As shown in Fig. \ref{compression4}, the SL-based FT scheme requires transmitting 144.76 GB of data to complete the fine-tuning process, which is 6.7 times that of our proposed scheme. Meanwhile, without compression, our scheme would require 14.75 times more data transmission, demonstrating the importance and effectiveness of the proposed compression scheme.
In Fig. \ref{compression3}, we present the gains achieved by different compression methods on transmission, reducing the final transmitted data to 6.8\% of the original activations, thereby effectively reducing communication overhead.

\begin{figure}[t]
\centering
\subfloat[CIFAR100 IID]{\includegraphics[width=1.6in]{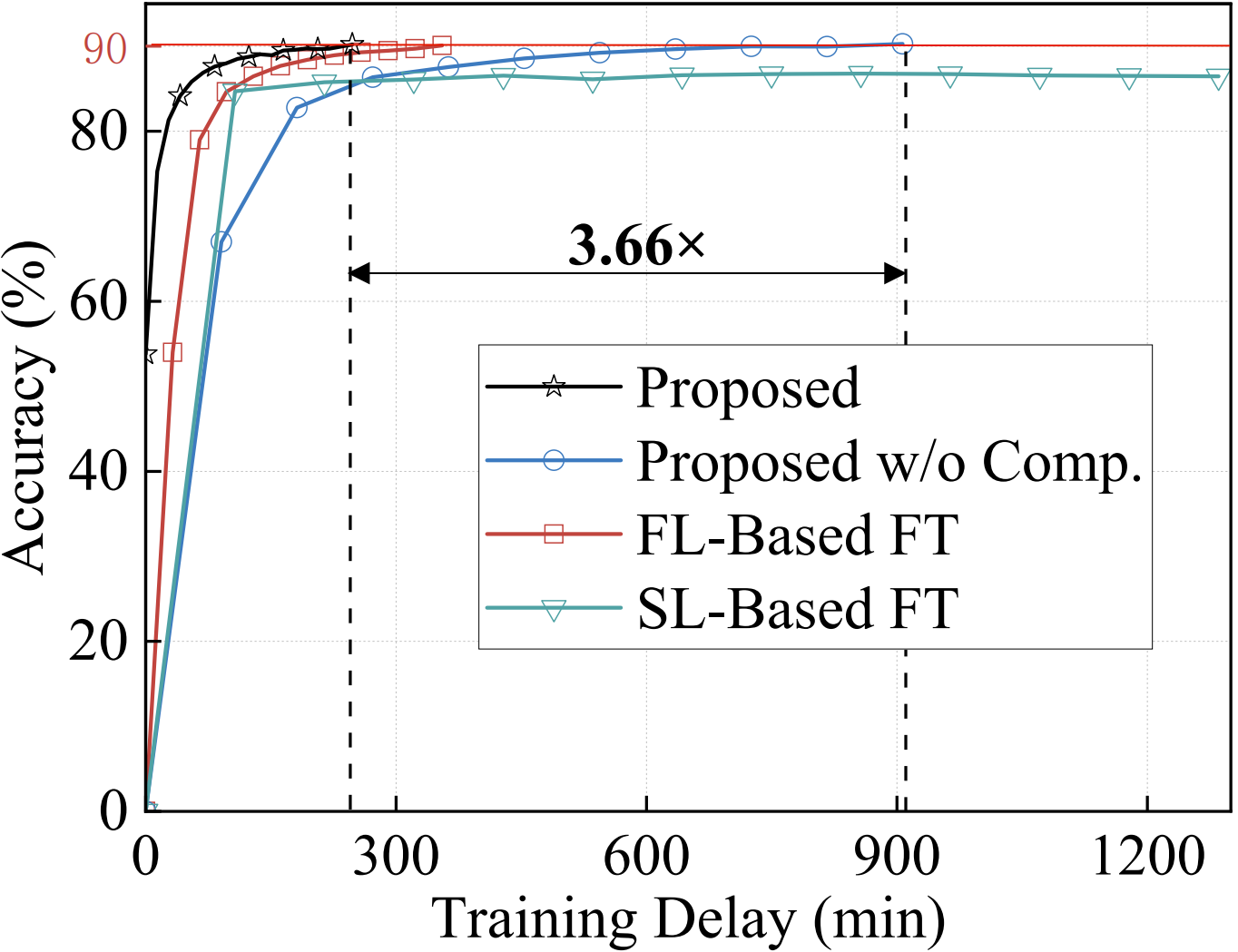}\label{cifar100_iid_delay}}
\subfloat[CIFAR100 non-IID]{\includegraphics[width=1.6in]{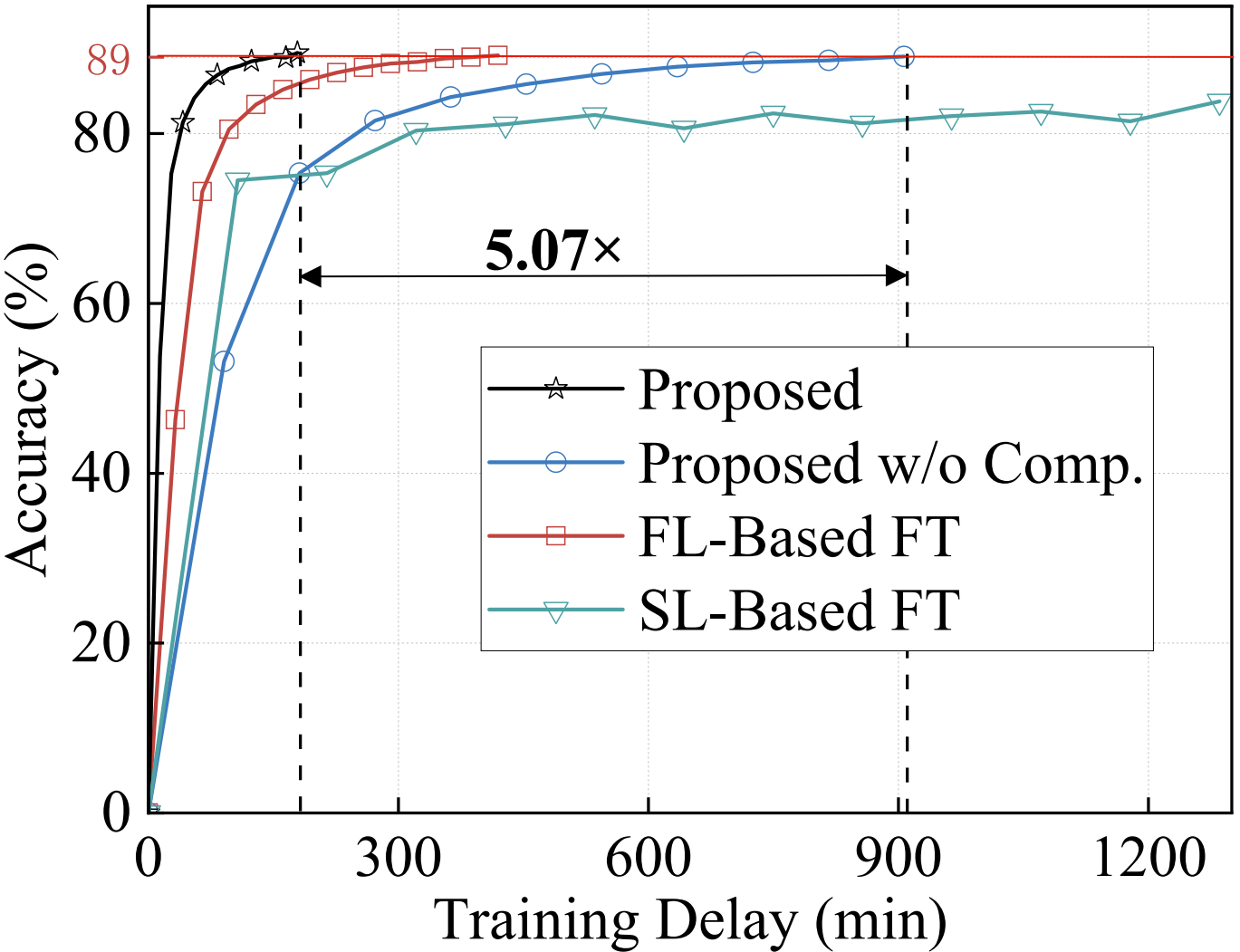}\label{cifar100_noniid_delay}}\\
\subfloat[Tiny-ImageNet IID]{\includegraphics[width=1.6in]{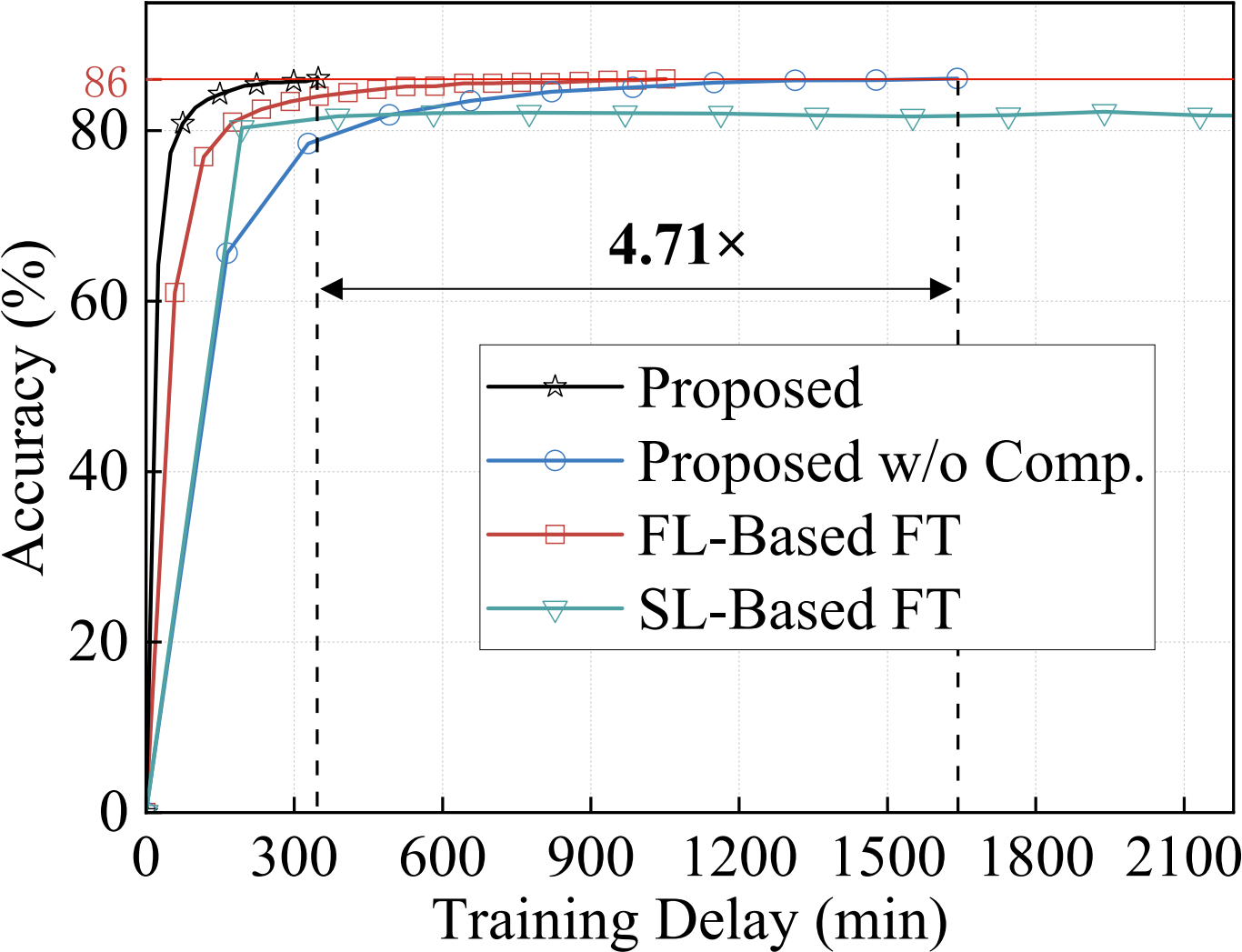}\label{tinyimagenet_iid_delay}}
\subfloat[Tiny-ImageNet non-IID]{\includegraphics[width=1.6in]{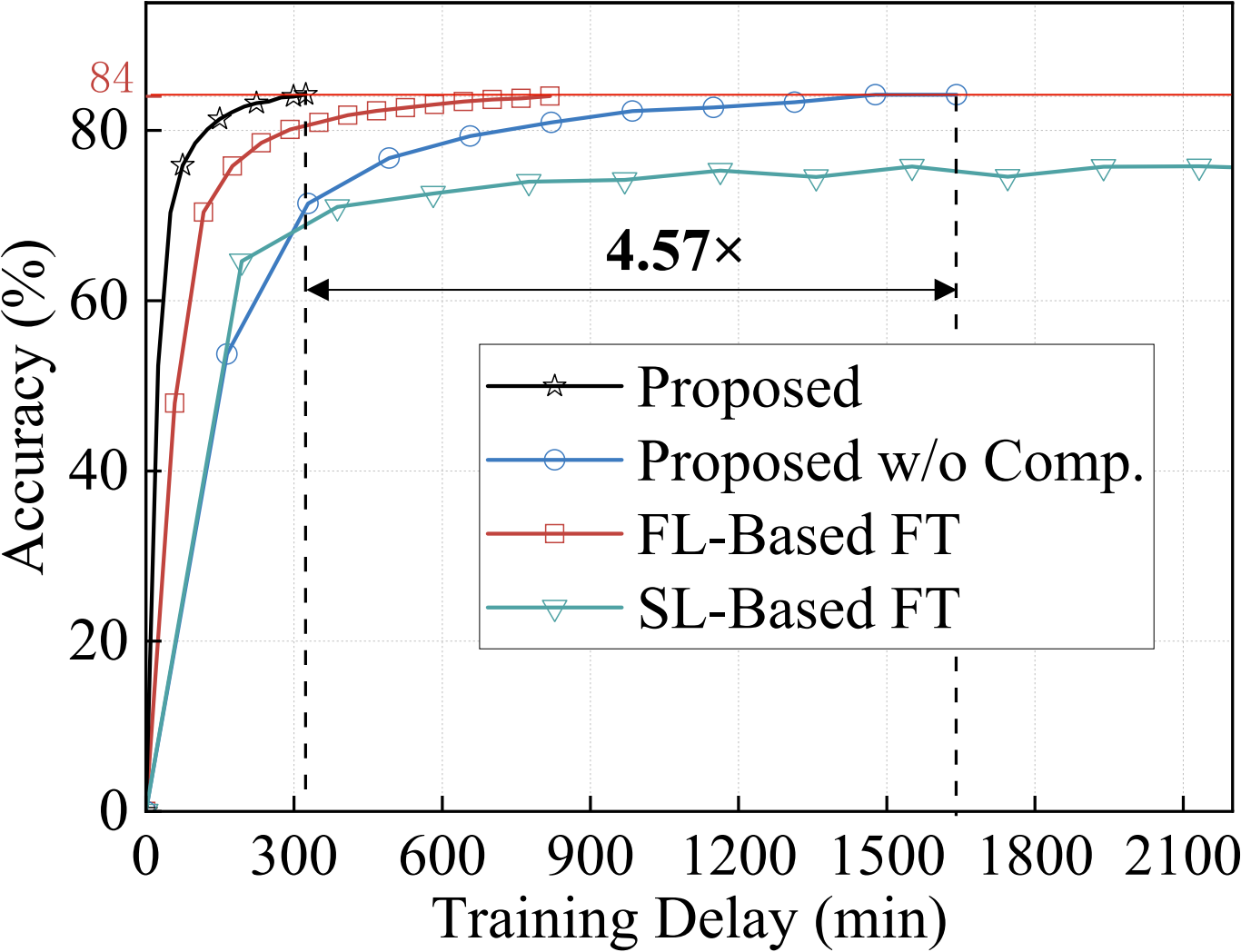}\label{tinyimagenet_noniid_delay}}
\caption{Delay performance of different fine-tuning schemes.}
\label{Delay2}
\end{figure}

\subsubsection{Fine-Tuning Delay}
Figure \ref{Delay1} illustrates the latency performance of our optimization algorithm compared to other resource allocation methods. We evaluated two baseline algorithms: one where bandwidth resources are evenly distributed among devices and another where bandwidth resources are randomly allocated. Experimental results demonstrate that under various system bandwidth conditions, the proposed two-stage optimization algorithm consistently outperforms the baselines. The proposed algorithm maintains its performance gains even in resource-constrained networks. For instance, even under a constrained 5 MHz bandwidth, the proposed algorithm reduces communication delay by up to 53.1\% in each fine-tuning round. This effectiveness is attributed to our communication compression scheme, which minimizes the communication volume to a very small scale during the fine-tuning process.

Figure \ref{Delay2} presents the delay performance results, evaluated using the time required to reach the highest accuracy of the FT scheme.  The results demonstrate that the proposed scheme achieves the fastest training speed across various datasets and under different data distribution conditions. For example, under the non-IID condition of CIFAR-100, the proposed algorithm requires only 178.8 minutes to reach the maximum accuracy. This is 2.34 times faster than FL-baed FT, which requires 419.6 minutes, and 5.07 times faster than the non-compressed version of the proposed scheme, which takes 906.9 minutes. Additionally, the proposed scheme is 6 times faster than the SL scheme, which reaches 88\% accuracy in 1071.1 minutes. The gain is attributed to the proposed scheme's integration of serial training and model aggregation, which ensures high accuracy, while the compression scheme significantly reduces the overall fine-tuning delay.

\section{Conclusion}\label{8}
In this paper, we propose a novel SFT scheme for LLM fine-tuning in wireless networks.
The SFT satisfies devices' memory constraints by splitting the LLM, accelerates fine-tuning with a parallelized SL scheme, and reduces communication overhead via a joint compression scheme.
Furthermore, we have developed a two-timescale resource management algorithm to minimize the overall fine-tuning delay. Experimental results demonstrate that the proposed scheme reduces fine-tuning delay and communication overhead while satisfying device-side memory and accuracy constraints. This scheme is well-suited for collaborative LLM fine-tuning across multiple memory-constrained devices in resource-limited wireless networks. 
In future work, we will investigate the performance of the proposed scheme in large-scale wireless networks.


\bibliographystyle{IEEEtran}

\bibliography{JSTSP.bib}

\vfill

\end{document}